# Second and third harmonic generation in topological insulator-based van der Waals metamaterials

Alessandra Di Gaspare[1], Sara Ghayeb,[1] Craig Knox[2], Ahmet Yagmur[3], Satoshi Sasaki[3], Mohammed Salih[2], Lianhe Li[2], Edmund H. Linfield[2], Joshua Freeman,[2] and Miriam S. Vitiello[1]

[1]*NEST, CNR-NANO and Scuola Normale Superiore, 56127, Pisa, Italy*
[2]*School of Electronic and Electrical Engineering, University of Leeds, Leeds, LS2 9JT, UK*
[2]*School of Physics and Astronomy, University of Leeds, Leeds, LS2 9JT, UK*

**Abstract**

**High-order harmonic generation (HHG) in solids – the frequency up-conversion of an optical signal – is governed by symmetries. At terahertz (THz) frequencies, HHG is a key technology to access high frequency spectral windows that are usually difficult to cover using conventional solid state laser technologies. This effect has been recently exploited in graphene where HHG has been demonstrated, albeit only at odd multiples of the driving frequency owing to its inherent centro-symmetry. In topological insulators (TIs), the combination of spin–orbit interaction and time-reversal symmetry create an insulating bulk state with an inverted band order, inseparably connected with conducting surface states. TIs have been predicted to support unconventional high harmonic generation from the bulk and topological surface, which are usually difficult to be distinguished. However, no experimental results have been provided, so far. Here, we exploit the strong optical field amplification provided by an array of single or double split ring resonators, with embedded $Bi_2Se_3$ or $(In_xBi_{(1-x)})_2Se_3$/ $Bi_2Se_3$ van der Waals heterostructures, to achieve up-conversion in the 6.4 (even) – 9.7 (odd) THz frequency range. This results from bulk centro-symmetry (odd states) and symmetry breaking in the topological surface states (odd and even).**

1. Introduction

Dirac materials (DM) possess gapless, linear energy band structures, and have massless Dirac carriers that dominate their optical and electronic properties at the few atomic monolayers' level[1,2,3]. They also display extraordinary nonlinear-optical properties[4], and graphene is in this respect the most peculiar example, owing to its Dirac carriers' electrodynamics. When light is absorbed in graphene, electron-electron interactions induce electron heating on a 10-100 fs timescale, which is followed by electron cooling, typically involving the emission of phonons on a picosecond timescale[5,6]. Over the past decade, the non-linear optical properties of graphene have been extensively investigated[7], and a wide range of associated mechanisms reported, such as saturable absorption[8,9], self-phase modulation[10], and four wave-mixing[11]. The ultrafast intraband dynamics[5,6], the centrosymmetric crystal nature of the honeycomb single layer graphene structure[12], and the inherent giant optical third-order nonlinearities[13,14] have also led to the demonstration of frequency up-conversion[13,14,15,16,17,18] – high harmonic generation (HHG) – in a broad spectral range from the ultraviolet[15] to the mid-IR[16], and down to the technologically-relevant terahertz (THz) frequency range, even at room temperature and in large-area synthetic films. This is a consequence of the third order non-linearity ($\chi^{(3)} \sim 10^{-9}$ m$^2$/V$^2$), that is almost seven orders of magnitude larger than that of typical semiconductor heterostructure lasers ($7\times \sim 10^{-16}$ m$^2$/V)$^{2,19}$, and significantly larger than that of alterntive materials exploited in THz photonics and electronics[20]. At THz frequencies, in DMs, the nonlinear effects responsible for the observation of HHG can usually be ascribed to the collective thermodynamic response of the free carriers to the driving field[21]. Crucially, unlike in other frequency domains, HHG can be induced at moderate fields ($\sim \leq 10$ kV/cm) using pump sources in the low THz frequency, and hence HHG does not require dedicated research facilities, such as the use of bulky laser systems[21] or accelerator-based sources[22]. Conversely, in the high $\geq 2$ THz frequency range, to overcome the small Drude weight of DM[23,24] – a parameter that quantifies the intraband absorption strength of the semi-metallic Dirac free carriers, and is limited by their number, $\sim 10^{-4}$ times lower than in noble metals[25] – either plasmonic effects or micro-resonators must be used to confine the THz fields to deeply subwavelength volume[26], and to reach electric field amplification factors up to two order of magnitude larger than in a bare film. Combining plasmonic confinement and resonant field enhancement in a non-linear DM[18] recently led to third HG in graphene using high power (>2 W) THz quantum cascade laser (QCL) as a pump[27].

Topological insulators (TIs)[32] have attracted considerable interest owing to their peculiar characteristic of having an insulating bulk state and highly conductive, topologically-protected surface states with a Dirac-like dispersion[28]. Dirac carriers in the topological surface states exhibit dissipation-free transport, owing to spin-momentum locking and suppressed back-scattering[29]. Furthermore, they have been proven to sustain robust long-lifetime collective surface excitations, i.e. plasmon-polaritons, from THz[30,31,32] to MIR[33,34] frequencies, thus enabling new applications in thermoelectric cryostats[35], magnetic storage devices[36], and nanophotonics, development of spintronic emitters[37] and topological quantum computing elements[38]. Second- and third harmonic generation (SHG and THG) have been recently reported in $Bi_2Se_3$, a widely studied TI[39,40]. Unlike graphene, SHG in TIs is not forbidden as $Bi_2Se_3$ has a non-zero second order susceptibility, stemming from inversion symmetry-breaking in the surfaces states[41] and phonon effects. Essentially, in $Bi_2Se_3$, with a centrosymmetric bulk, the second-order response of THz-SHG can arise only when the optical response is dominated by the topological surface states[42,43,44]. However, in a recent paper[40], demonstrating THG in the THz range in a $Bi_2Se_3$ film, SHG was not observed. The absence of SHG usually arise when there is a significant interplay between surface and bulk states, e.g. when the Coulomb interaction with the bulk carriers becomes the most proficient path for the ultrafast heating-cooling dynamics controlling the nonlinear response[40]. In an alternative picture[39], the suppression of second-order processes was ascribed to uneven population in the topological surface states in twinned crystal domains[39]. This highlights the importance of developing sophisticated and systematic techniques to synthetize large-area and scalable TI films, with accurate control of the orientation and the quality of the crystal growth, particularly when surface symmetry breaking is sought. Recently, molecular beam epitaxy (MBE) has emerged as a powerful method for growing $Bi_2Se_3$ films of controlled thickness and composition, [45,46,30,47] and for devising tunable topological surface states through atomically controlled multilayering of different TI compunds[48,46]. As an example, the deposition of an insulator ultrathin film adjacent to the TI surface[46] was revealed to be a promising strategy for achieving a robust insulating bulk with surface states possessing a tunable Dirac point. Indeed, the incorporation of a thin Sb layer in $Bi_2Te_3$, allowed for the tuning of the energy of the Dirac point[48] in the topological surface state formed in the $Bi_2Te_3$ [48].

In this work, we exploit the strong optical field amplification provided by a metamaterial array of single or double split ring resonators, embedding $Bi_2Se_3$ film or a set of quantum designed $In_xBi_{(1-x)})_2Se_3/Bi_2$-based van der Walls heterostructures having engineered thicknesses, both grown by molecular beam epitaxy, to innovatively tailor, either odd (governed by bulk centro-symmetry) or even and odd orders frequency up-conversion in the THz frequency range, induced by symmetry breaking in the topological surface states. This is achieved by optically pumping the resonator array with a high-power THz QCL. The mode confinement and field enhancement in the split gap/gaps, combined with the high-power density of the QCL ($\sim \leq 1$ kW/cm$^2$), is essential to activate THG and SHG in the largely unexplored high THz (>3 THz) frequency range. The thin $(In_xBi_{(1-x)})_2Se_3$ virtual substrate mitigates the formation of defects at the substrate-TI interface, which can lead to the formation of trivial interface states. This increases the proportion of carriers in the topological surface state, and improves its overall quality, leading to SHG, a phenomenon not allowed in $Bi_2Se_3$ sample since the interaction of Dirac and bulk carriers prevents the symmetry breaking needed to allow the second-order optical process, even if the TI surfaces remain protected[40].

## 2. Results

### 2.1 $Bi_2Se_3$ spit ring resonator: Design, realization and characterization

We initially engineer an array of micrometric circularly shaped single-split ring resonators (SSRR) on a THz transparent $SiO_2/Si$ substrate. The resonator geometry was matched to the frequency of the impinging high-power QCL, which has an emission centered at 3.25 THz, to exploit the field enhancement in the split gap of each individual SSRR[49,50], following a design optimization procedure presented elsewhere[27,49]. The integration of the $Bi_2Se_3$ in the resonator array requires an accurate model that accounts for the complex optical constants of the film, to evaluate the field enhancement and the optical response. Here, to mitigate the optical screening of the metallic SSRR array, and the consequent resonance bleaching associated with the semi-metallic character of the Dirac carriers in the TIs, we integrate the $Bi_2Se_3$ only in the SSRR split gap. The large area ~20 nm thick $Bi_2Se_3$ layer described adopted in the present work in described in ref.[30]. The near field optical mapping of this film shows that the material sustains propagation of Dirac plasmon polaritons[30], the dispersion of which is dominated by the Dirac carriers hosted in the surface topological linear bands, the 2D massive electrons in the trivial surface bands, and, the dielectric response of the thin bulk layer[30,31]. Hence, in agreement with the methods of refs.[30,31,32], its total conductivity is given by the sum of the bulk conductivity ($\sigma_{bulk}$), the massive surface carrier conductivity ($\sigma_{2DEG}$), arising from the Shockley-type band-bending at the surface[30], and the conductivity of the Dirac like states ($\sigma_{DM}$).

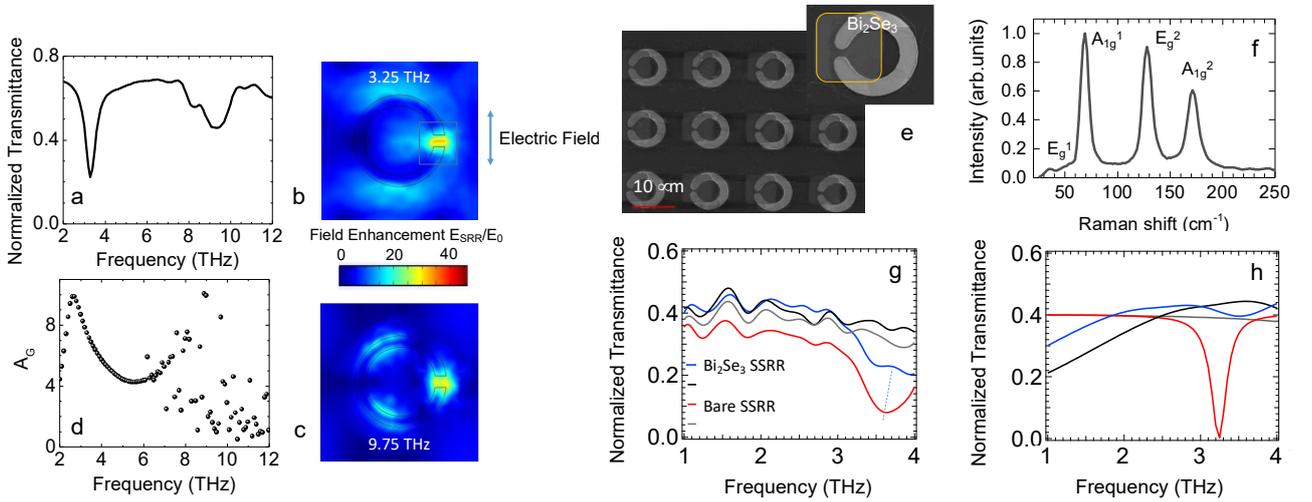

**Figure 1 (a)** Simulated transmittance, calculated using a finite element method (FEM), of the $Bi_2Se_3$-SSRR, for an electric field polarization parallel to the dipolar split gap; **(b-c)** two-dimensional (2D) profiles of the electric field ($E_{SRR}$) calculated using FEM simulations of the $Bi_2Se_3$-SSRR for the (b) fundamental and (c) third harmonic frequency modes, normalized by the field on the surface outside the resonator, $E_0$ (zero enhancement region). **(d)** Amplification ($A_G$) factor as a function of frequency calculated for the $Bi_2Se_3$-SSRR, defined as the ratio of the average electric field on the split ring resonator plane to the average electric field impinging on a $SiO_2$/Si substrate under the same boundary conditions and input settings in the simulator module. **(e)** Scanning electron microscope image of the fabricated $Bi_2Se_3$ embedded SSRR array. The yellow areas (false color) highlight the patterned $Bi_2Se_3$ **(f)** Micro-Raman spectra measured on the $Bi_2Se_3$ SSRR, after transfer and processing. **(g)** Normalized transmittance of the $Bi_2Se_3$-integrated (blue, black curves) and bare (red, gray) SSRR measured using time domain spectroscopy (TDS, Terascan by Menlo Systems), in a nitrogen-purged environment, with the dipole of the SSRR oriented parallel (blue, red) and perpendicular (black, gray) to the linearly polarized beam of the TDS transmitter. Each transmittance curve was obtained by Fourier transformation of the time-domain traces, normalized by the transmission through the $SiO_2$/Si substrate. The dotted line marks the blue shift of the blue curve **(h)** Calculated transmittance (COMSOL Multiphysics), with the same color code as panel (g).

We then calculate the total transmittance (Fig. 1a) of the $Bi_2Se_3$/SSRR using finite element method (FEM) electromagnetic simulations (COMSOL Inc.), by assigning the total optical conductivity (see (See Supplementary Information (SI), section S1), as a transition boundary condition to a 5×5 μm$^2$ rectangular area around the gap (see Fig.1b). The $Bi_2Se_3$/SSRR calculated transmittance shows a resonant absorption dip (Fig.1a) matching the optical pump emission at $v_o$ = 3.25 THz, and a quality factor $Q$ = 5.1±0.2 for the present design, with an additional broader dip located at $3v_o$.

The 2D map of the field enhancement of the SSRR (amplification factor $A_G$) at the resonance, is reported in Fig.1b. Here $A_G$ is defined as the ratio between the average electric field on the TI-embedded SSRR surface to the average electric field from a plane wave illuminating an unpatterned surface (i.e. a $SiO_2$/Si substrate), assuming the same incoming power and boundary conditions on the input port in the simulation. In the gap area, $A_G$ is more than one order of magnitude larger than the absorption in an unpatterned film. This leads to the formation of an intense hotspot in the active film,[51,52] owing to the dipolar resonant mode sustained by the two arms of the split gap, that maximizes the light–matter interaction, which, in turn, shifts and enhances the nonlinear terms of the $Bi_2Se_3$ optical conductivity at the QCL pump frequency, increasing the HHG conversion efficiencies[27]. Interestingly, the SSRR has resonant modes even at harmonics higher than the fundamental, as shown in Fig.1c for the third harmonic, boosting the overall THG efficiency[14]. The presence of a 3$^{rd}$ order resonance in the transmission, having a quality factor $Q_3$=5.3±0.2, is reflected in the frequency dependence of the field enhancement (see Fig.1d), which in fact shows a second frequency range of amplification, also around the 3$^{rd}$ harmonic frequency range.

We then fabricated the designed $Bi_2Se_3$/SSRR structure. We first realized the metallic SSRR array, of size 5×5 mm$^2$, on a $SiO_2$/Si substrate, using direct laser-write photolithography, followed by deposition and lift-off of a Cr (5 nm) /Au (100 nm) metal bilayer. Next, the MBE $Bi_2Se_3$ film was transferred on top of the patterned sample, using a PMMA assisted wet transfer method[53]. Finally, a second optical lithography step was performed to define and mask the gap region area, and then to remove the $Bi_2Se_3$ film from all areas surrounding the gap (Fig.1e). To assess the quality of the $Bi_2Se_3$ film after the transfer and processing, we performed micro-Raman spectroscopy directly on the $Bi_2Se_3$ integrated into the resonator (Fig.1f). Raman spectra (Horiba, Explora Plus microscope) were measured using a 532 nm laser delivering 1 mW optical power, focused to a ~2μm spot. Three visible peaks associated with the three prominent vibrational modes of $Bi_2Se_3$, namely $A_{1g}^1$, $A_{1g}^2$ and $E_g^2$, are retrieved at 69.0, 171.8 and 128.5 cm$^{-1}$ wavenumbers, respectively, with a weaker peak at 37.1 cm$^{-1}$ assigned to the $E_g^1$ mode; this is in agreement with previous characterizations of the as-grown MBE wafer[30], confirming that the good crystalline quality of the MBE synthetized films was not affected by the processing.

The transmittance (Fig.1g), measured using time domain spectroscopy (TDS), shows a resonant response, visible only when the axis of the dipolar resonance of the ring is oriented parallel to the polarization direction of the probe beam, thus proving the resonator sensitivity to the incoming field polarization. When we compare the transmittance of the $Bi_2Se_3$-integrated SSRR array with that retrieved from the pristine resonator array having an identical design, we observe a slightly weaker (~22%) and blue-shifted (0.20 THz) resonance. Both frequency shift and resonance broadening arise from the optical response of the $Bi_2Se_3$ film. Indeed, the semimetal character and the inductive response in the 2D material embedded SSR result in a "screening" effect of the metallic resonator, with the consequent resonance weakened by the absorption occurring in the $Bi_2Se_3$ material.

The frequency shift and resonance broadening of the optical response is clearly visible in Fig.1h, which shows the simulated transmittances, in broad agreement with the experimental data (Fig.1g). Notably, the field enhancement is also slightly affected by the presence of the TI film, in both its frequency behavior and magnitude (see SI, S2). This proves, as expected, that the resonator acts as a field enhancer, independently of the active film's presence, indicating that the optical amplification of the resonator can be reliably assessed in both the pristine and the TI-integrated cases. However, in order to match the desired pump frequency and assess accurately the real electric field magnitude on the nonlinear film, the full complex conductivity must be included in the resonator design.

**2.2. Third harmonic generation (THG) in externally pumped $Bi_2Se_3$/SSRR array**

To demonstrate THG in the $Bi_2Se_3$ embedded resonator, we performed a pump-probe experiment (Fig.2a), in which the spectral content of the emission from the $Bi_2Se_3$ SSRR array was probed under intense (~1 kW/cm$^2$) pump beam illumination.

Signals emitted by the $Bi_2Se_3$ -SSRR were acquired in step-scan mode, under vacuum and with a resolution of 1cm$^{-1}$, after filtering the fundamental frequency of the QCL, with the 2-mm-thick high-bandpass Ta crystal filter. We first oriented the dipolar mode axis of the SSRR along the polarization axis of the QCL, then perpendicular. A peak at ~9.75 THz, signature of THG, at the 3$^{rd}$ order harmonic of the pump beam (Fig.2c, green and blue curves, acquired on two different samples from separated fabrication runs) was only measured when the polarization of the pump beam coincides with the dipolar orientation of the SSRR, illustrated in Fig.1b. No evidence of possible signals at the 3$^{rd}$ (or 2$^{nd}$) harmonic frequency was retrieved when the resonator was pumped along the orthogonal direction (cross-polarization pumping). Although ≥ 7 hours long scans were required to acquire data for both samples, a consistent intensity was measured for the THG signal ~ (1.3 ± 0.3)×10$^{-5}$ of the pump signal; this was determined after accounting for the acquisition conditions (including the gain conversion of the lock-in amplifier and FTIR aperture diameter). This value compares with a noise-level that is ≤ 0.5×10⁻⁵ the intensity of the pumping beam, i.e. the measured THG is not below the instrumental limits. The retrieved THG signal is ~10⁻⁵ less intense than the most intense mode of the pumping QCL, and it is comparable with the conversion efficiency reported in single layer graphene[27]. No signature of SHG was observed. This is in agreement with previous reports [40], showing only odd-harmonic generation in $Bi_2Se_3$; indeed, it is theoretically predicted[39] that SHG would require symmetry breaking in the topological surface state. This reflects a topological protection strong enough to guarantee the full isolation of the optical response of the surface states from those in the bulk[39], i.e. the surface states are not affected by the bulk. Specifically, the TI surface state in the present sample, with a mobility measured on the as-grown film of 1500±100 cm²/Vs[30], was shown to be strongly affected by the presence of lower mobility carriers in the trivial surface state, i.e. the massive 2DEG, and by the bulk conductivity – this was revealed by the absence of Landau Level quantization in low-T magneto-transport, up to an 8 T magnetic field[30].

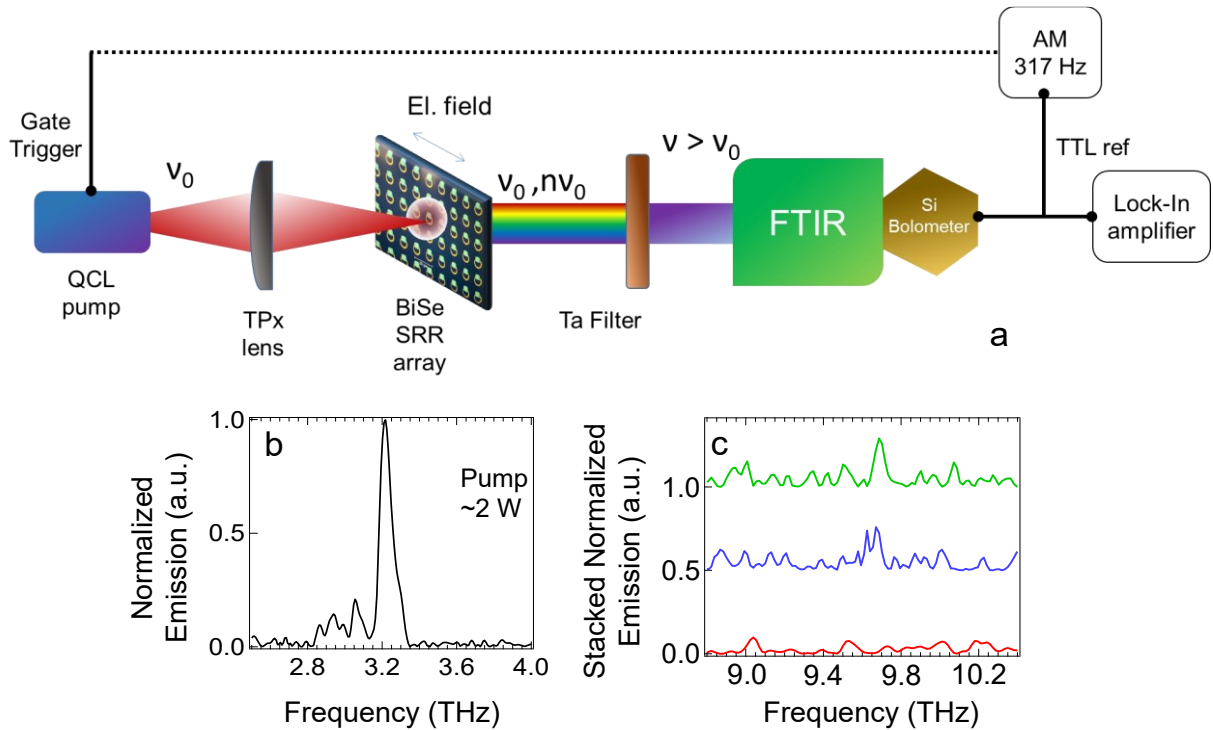

**Figure 2: (a)** Schematic diagram of the experiment apparatus for investigating SHG and THG. Light from a single-plasmon QCL, emitting 2.5 W into free space, is focused on the resonator array. The SHG and THG signal is isolated after filtering the THz QCL with a 6 THz high-bandpass Ta-filter. **(b)** Fourier transform infrared (FTIR) emission spectrum of the QCL pump source, acquired under vacuum with a 1 cm⁻¹ spectral resolution. **(c)** Emission spectrum at the third harmonic measured on the optically pumped $Bi_2Se_3$-SSRR from two different fabrication runs (green and blue), after filtering the QCL with the Ta-filter. Spectra were measured in step-scan mode with a spectral resolution ~1cm⁻¹ and aperture size~5mm, with the split-gap dipole of the SRRs in the array oriented parallel (green and blue) and perpendicular (red) to the laser linear polarization direction.

## 2.3. Third harmonic generation: modeling of the optical response in the Bi$_2$Se$_3$ resonators

Since the investigated Bi$_2$Se$_3$ samples did not show any evidence of SHG, we restrict our modeling to third order processes. We calculate the conversion efficiency by implementing the same analytical model developed for graphene-SSRR arrays[27]. We assume that the non-linear optical response of the Bi$_2$Se$_3$ film stems exclusively from the hot-electron dynamics in the massless Dirac carriers in the topological surface states[40,7]. The bulk affects the overall resonator response, as shown by the resonance shift/broadening and by the modulation of the field enhancement. However, it has also been recently reported[40] that the bulk carriers provide an alternative ultrafast dissipation channel for the hot electronic temperature, through surface-bulk Coulomb interactions. This property induces an increase in the up-converted signal amplitude of more than two orders of magnitude[40] compared to widely-studied graphene samples[54].

In the external pump scheme employed in the present work, the THG conversion efficiency (CE) is defined as the variation of the sample transmission normalized by the linear transmission, $CE = (T_3 - T_0)/T_0$, and is evaluated through the Tinkham formula[55], as follows:

$$[T_0, T_3(\nu)] = \frac{4n_{sub}}{Z_0^2 \left| \frac{n_{sub}}{Z_0} + [\sigma_0, \sigma_{3,tot}(\nu)] \right|^2} \tag{1}$$

where $Z_0$ is the vacuum impedance and $n_{sub}$ is the refractive index of the substrate. Following eq.1, CE is then related to both the linear ($\sigma_0 = \sigma_{DC} + \sigma_{2DEG} + \sigma_{Bulk}$) (see SI) and the field-driven nonlinear ($\sigma_{3tot} = \sigma_0 + |E_0(\nu)|^2 \times \sigma_3(\nu)$) optical conductivities. In the latter expression, $\sigma_3(\nu)$ is the non-linear Kerr optical conductivity of the BiSe topological carriers, calculated accounting for the hot-electron temperature (T$_e$) dependence of the Drude coefficients[56,57], and including the contribution of the plasmonic frequency of the resonator array by substituting the frequency ν with the effective spectral frequency $\nu_{eff} = (\nu^2 - \nu_0^2)/\nu$ (see SI, section S3). To mimic the operation of our external pump, comprising a QCL driven in pulsed mode at ~kHz repetition rate and with pulse duration ~μs, T$_e$ is evaluated for steady state illumination [57,27,54], i.e. when the optical beam is either continuous wave or with a pulse duration much longer than the cooling time (~ps). Under this assumption, T$_e$ depends on the ratio between the incoming pump power, the Bi$_2$Se$_3$ cooling time of the TI film (τ$_{cool}$~3.5 ps[58]), and the heat capacitance of the hot fermions of the Bi$_2$ Se$_3$ (C$_{BiSe}$~ 0.16 μWm$^{-2}$sK$^{-1}$). Remarkably, unlike the case of few-ps pulsed pumping[59,5,40], in these condition, namely with ~10$^7$ THz cycles driving the hot carrier populations, a slower cooling channel is not detrimental for efficient THG, but rather helps to build up higher T$_e$, with a consequent stronger modulation of the T$_e$-dependent conductivity, which in turn leads to a larger harmonic generation efficiency.

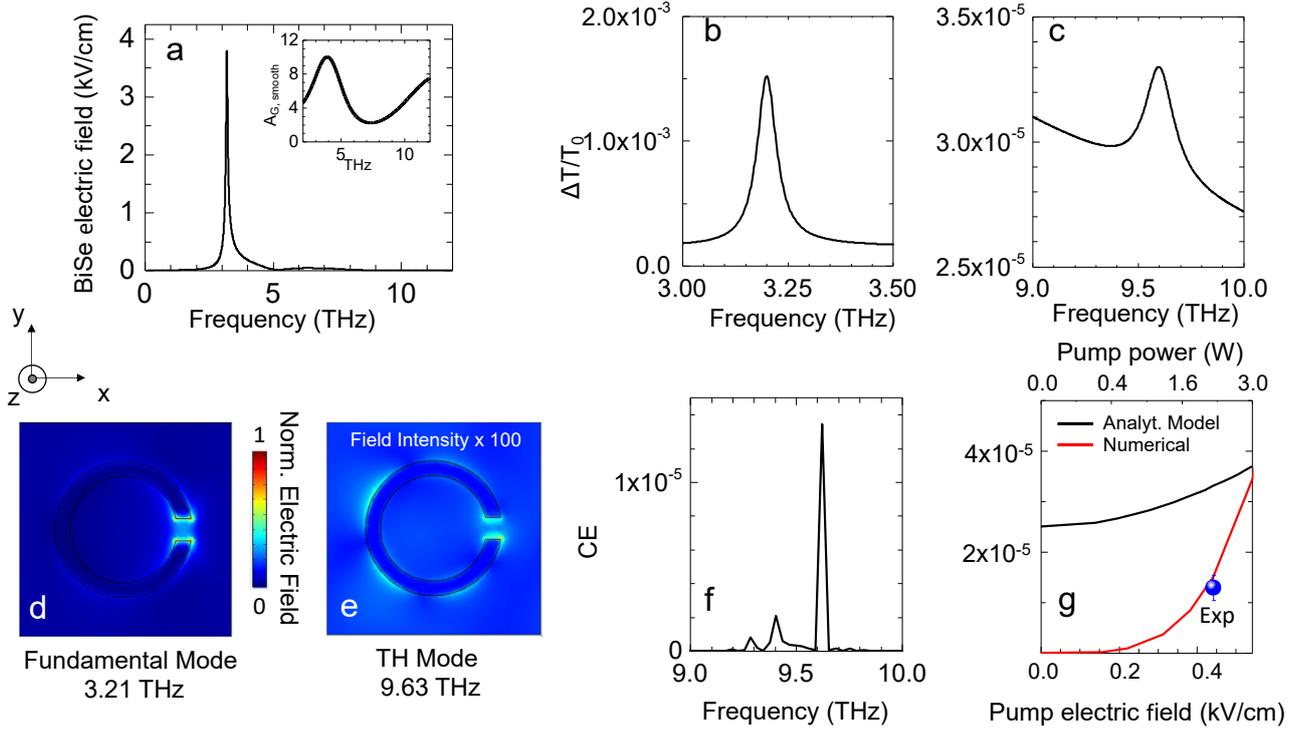

**Figure 3: (a)** Electric field incident on the $Bi_2Se_3$ SSRR as a function of frequency, calculated through the convolution of the field enhancement shown in the inset, and the optical beam intensity modeled assuming a 2 W Gaussian-peak emission power focused on a 700 μm diameter circular spot. **(b-c)** $\Delta T/T_o$ as a function of frequency in the pumped $Bi_2Se_3$ SSRR, assuming the incident effective electric field from (a), and following the analytical model illustrated in the text, plotted (b) at the fundamental QCL frequency $\nu_0$, and (c) at $3\nu_0$(c). **(d-e)** Two-dimensional (2D) profiles of the emitted electric field from the $Bi_2Se_3$ SSRR, calculated according to the refined numerical model illustrated in the main text at (d) the fundamental resonance frequency (3.21 THz), and (b) the third harmonic frequency, 9.63 THz. **(f)** Third harmonic conversion efficiency (CE) as a function of frequency calculated using the surface current density numerical model, assuming an input power of 2W. **(g)** THG efficiency, $\Delta T/T$ ($3\nu_0$), in the $Bi_2Se_3$ SSRR, as a function of impinging electric field (bottom axis) and total power (top axis), calculated assuming an incident beam diameter of 700 μm, with simplified analytical (black) and refined numerical (red) models, respectively. The blue circle represents the CE value retrieved from the experiment.

Figure 3a shows $|E_0(\nu)|^2$, the effective electric field driving the nonlinear response. It is given by the convolution, as function of frequency, of the field enhancement, numerically extracted from the FEM simulation data in Fig.1d both at the fundamental and at the third harmonic frequency (and plotted in the inset of Fig.3a using a 10-point, adjacent-averaging smoothing procedure), and the optical electric field of the external beam, modelled as a quasi-monochromatic radiation field delivering 2 W optical power, focused on a 0.4 mm diameter circular spot. The effective electric field in Fig.3a was used to calculate the third harmonic CE as a function of frequency (Fig.3b-c).

Owing to the photo-induced $T_e$ dependance of the conductivity, the calculated transmission variation $\Delta T/T_0$ at the fundamental frequency $\nu = \nu_0$ increases (Fig.3b), a phenomenon ascribed to the absorption saturation already reported in DM[8], and specifically in TIs[60,61]. In addition, another peak-like increase is predicted in $\Delta T/T_0$ at $\nu = 3\nu_0$, corresponding to a CE$\sim 3.3\times 10^{-5}$ for THG emission (Fig.3c); this is the only high-order harmonic term included in our model, and is also boosted by the field enhancement in the CSSR. Noticeably, the frequency curve $\Delta T/T_0$ also shows a baseline of $\sim 2\times 10^{-5}$ around the third harmonic peak. This arises from the non-zero values of the nonlinear optical conductivity, at all frequencies $\nu \neq 3\nu_0$, which lead to small, non-zero background contributions to $\Delta T/T_0$.

To better account for the role of the field amplification of the SSRR to the THG efficiency, we implemented an alternative, more sophisticated, numerical model to predict the THG process in the plasmonic structure, by following the approach proposed in [62]. This method allows one to extract the THG efficiency as an output parameter directly from the simulation, by setting up the equations for the third harmonic generated field in the software module. The simulation layout comprises the same unit cell for the resonator, as used in the linear simulation of Fig 1. The method is then based on the assumption that the $Bi_2Se_3$ TI active film is a nonlinear surface current generator, defined as $J = \sigma_0 E_{TH} + \sigma_3 E_{FH}$, where $E_{FH}$ and $E_{TH}$ are the electric fields induced at the fundamental and third harmonic, respectively, with $\sigma_0$ being the total $Bi_2Se_3$ conductivity, and $\sigma_3$ the 3$^{rd}$ order non-linear conductivity, again expressed as the $T_e$-dependent Kerr nonlinear term[63] (see SI).

Within this method, the electromagnetic simulator can compute the CE of THG, by solving the nonlinear Maxwell equations at the third harmonic frequency. By exciting the resonator structure externally (from the top), it can be verified that it possesses the desired resonant mode (Fig.3d)). This result is implemented by using an electromagnetic input port, for which the power density – the most critical parameter to estimate the conversion efficiency – is a freely defined parameter. To extract quantitatively the CE, defined as the ratio between the delivered power at the third harmonic and the input power, the input port power is set to match the power density range used to illuminate the $Bi_2Se_3$-SSRR. The electric field profile at the third harmonic frequency (Fig.3e) shows a distribution like the fundamental mode, although is characterized by a field intensity more than two order of magnitude lower. The calculated frequency dependence of the CE (Fig.3f) well agrees with the field-enhancement driven emission at the third harmonic frequency. The CE value predicted at the resonance is $\sim 10^{-5}$, in agreement with the experimental results, providing the improved accuracy of the adopted mythology.

**2.4. Second and third harmonic generation in $(In_xBi_{(1-x)})_2Se_3/ Bi_2Se_3$ heterostructures**

The absence of SHG in the $Bi_2Se_3$ SSRR may be ascribed to the centrosymmetric nature of the $Bi_2Se_3$ bulk crystal dominating the optical response, combined with a lack of symmetry breaking in the topological surface state; more trivially, being a nonlinear mechanism, it could also be an effect of an inadequate electromagnetic coupling with the pump field. We thus investigated a set of $Bi_2Se_3$- heterostructure, comprising the incorporation of quintuple layers (QLs) of the trivial insulator $(In_xBi_{(1-x)})_2Se_3$ adjacent to the $Bi_2Se_3$ slab[64]. In the first heterostructure sample (H1), a 20 nm thick $Bi_2Se_3$ layer was grown onto a 30 nm thick insulating $(In_xBi_{(1-x)})_2Se_3$ buffer layer, itself deposited onto a sapphire substrate. In a second sample (H2), the same $(In_xBi_{(1-x)})_2Se_3$ layer was grown on both sides of the $Bi_2Se_3$, resulting in an $(In_xBi_{(1-x)})_2Se_3/Bi_2Se_3/(In_xBi_{(1-x)})_2Se_3$ three-layered structure, a prototypical building block for van der Waals (vdW) heterostructures, that could be repeated to give thicker, stacked MBE growth.

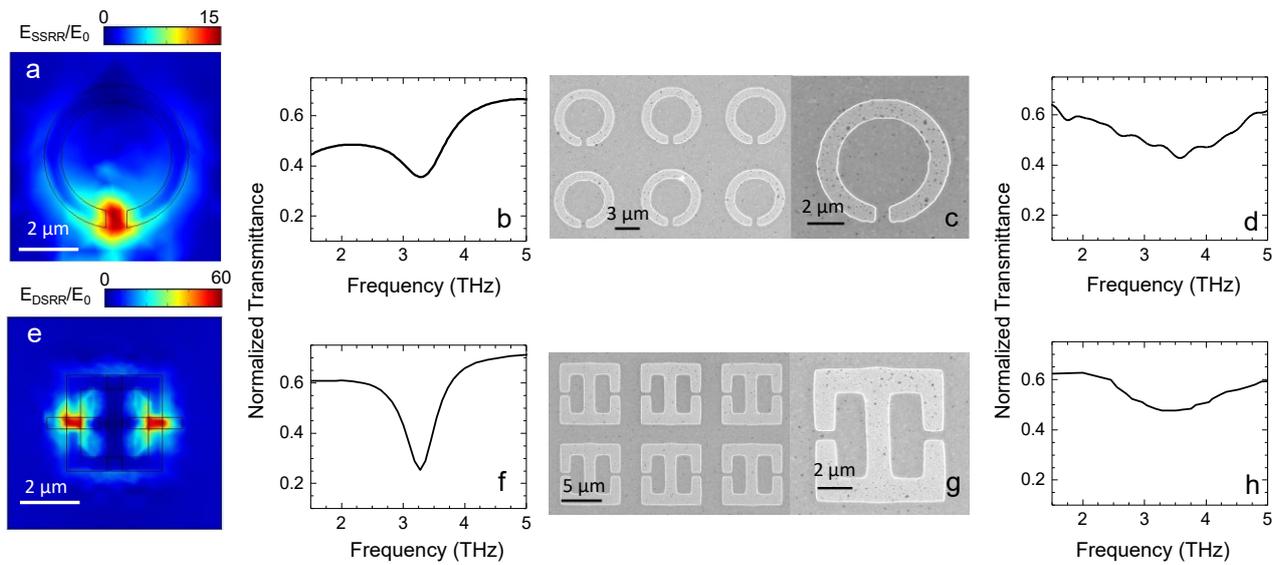

**Figure 4. (a)** 2D map of the normalized electric field ($E_{SSRR}/E_0$) on the $Bi_2Se_3$ SSRR at the resonance frequency $\nu_{SIM}$~3.245 THz, calculated using FEM simulations on an SSRR fabricated on as-grown $Bi_2Se_3$ heterostructures. $E_0$ is the field on the bare sapphire surface. **(b)** Simulated optical transmittance of the design shown in (a), extracted when the input beam polarization is oriented parallel to the split gap dipole. **(c)** SEM images acquired on the SSRR fabricated on the $(In_xBi_{(1-x)})_2Se_3/Bi_2Se_3/(In_xBi_{(1-x)})_2Se_3$ sample, H2. **(d)** Experimental transmittance acquired on the sample shown in (c), using FTIR, under vacuum, with a spectral resolution of 1 cm$^{-1}$, the internal MIR (Globar) source and a helium-cooled Si bolometer. A rotating, wire-grid linear polarizer was positioned in front of the sample; the reported transmittance curve was obtained by taking the ratio of the curve acquired by selecting the linear polarization parallel to the split gap dipole (polarization angle 0°), with the one acquired at 90°, and then normalizing to account for the reflection losses of the sapphire substrate (~40%). **(e)** Normalized electric field ($E_{DSRR}/E_0$) map of the $Bi_2Se_3$ DSRR at the resonance frequency $\nu_{SIM}$~3.25 THz, calculated using FEM simulations for the array realized directly on the as-grown MBE $Bi_2Se_3$ sample. **(f)** Simulated optical transmittance of the design shown in (e), extracted when the input beam polarization is oriented parallel to the split gap's dipole. **(g)** SEM images acquired on the DSRR fabricated on the $(In_xBi_{(1-x)})_2Se_3/Bi_2Se_3/(In_xBi_{(1-x)})_2Se_3$ sample. **(h)** Experimental transmittance acquired on the sample shown in (g), using the same methods as (d).

Micro-Raman spectroscopy experiments were performed to characterize the materials (see SI, section S7).[65] To shed light on the topological nature of the heterostructure-TIs, we also conducted low-temperature magneto-transport measurements. Both samples show resistances that decrease on cooling, indicating that scattering within these samples is dominated by electron-phonon interactions at high temperature (see SI, section S8). The low temperature magnetoresistance also shows weak anti-localization at low field[65], followed by a parabolic background in the longitudinal resistance which is indicative of ordinary magnetoresistance[65]. Furthermore, there is some slight non-linearity to the Hall coefficient, indicating the presence of multiple carrier species. The retrieved CE $\sim 10^{-5}$ for THG on the single $Bi_2Se_3$ sample (Fig.3) approached the signal-to-noise level in the pump and probe experiment. From the expected behavior of topological states in TIs, the 2$^{nd}$ order nonlinearity, $\chi^{(2)}$, $\leq \chi^{(3)}$ at best, and so similar CE values are expected[39] even for SHG in symmetry-breaking, pristine TIs. We, therefore, conceived an alternative resonator design to maximize the electric field amplification. This comprised a double split ring resonator (DSRR)[66], realized by integrating, back to back, two split rings in one single resonator. A comparison between the single-, already tested, and the new double-SRR is shown in Fig. 4. To minimize any possible detrimental effects on the material quality associated with handling and processing, we opted for a fabrication approach that preserve the properties of the TI surface. Specifically, we patterned the resonator array directly on the as-grown MBE TI films, H1 and H2, and restricted the fabrication to one single step, i.e. without performing the etching step for the isolation of the active region around the split-gap amplification area (see schematics in Figs 5a-b). First, we adjusted the resonator geometry (dimensions) for fabrication on a sapphire MBE substrate, while matching the pump frequency. During this design optimization, we adapted the model for the $Bi_2Se_3$ optical constants, discussed in Section 2, to match the conductivity of the heterostructure extracted from the magneto-transport experiment. The resulting SSRR design (see Fig.4a) has an overall smaller size, accounting for the higher real part of the sapphire refractive index $n_{Al2O3} \sim 3.1$ when compared silicon oxide, $n_{SiO2} \sim 2$. More specifically, the sapphire substrate was modeled as a dielectric layer with non-zero real and imaginary optical constants, to account for the absorption in the THz and MIR ranges (see SI, section S4). At resonance, the field enhancement in the split gap area of the SSRR is comparable to the SSRR on $SiO_2$ (see Fig.4a), as expected in a simply scaled geometry, and the calculated transmittance shows a resonant absorption centered at $\nu_{SIM} \sim 3.25$ THz and a $Q_{SIM} \sim 3.3$ (see Fig.4b). The transmittance measured on the sample H2, after fabricating the SSRR resonator array using the scaled design (Fig.4c), shows a resonance at a slightly higher frequency, $\nu_{EXP} \sim 3.52$ THz, and a significantly broader width, corresponding to a $Q_{EXP} \sim 1.53$ (see Fig.4d). The discrepancy in the resonant frequency may arise from the fabrication process, such as a thicker metal, implies a $\sim 0.3$ THz mismatch between the resonator mode and the pump laser line, that here is partially compensated by the broader FWHM. However, the higher resonance width is not necessarily detrimental in the THG process, since the field enhancement is only weakly dependent by the Q-factor [27].

We then tune the geometry and assess the field enhancement in the DSRR (Fig.4e,f). The field enhancement, extracted for input beam polarization oriented parallel to the axis of the split-gap dipolar axis, reaches values of the order of $\sim 10^2$ (Fig.4e). The field amplification region is also extended well behind the two split-gap areas. This behavior, together with the evidence that the electromagnetic field is mostly concentrated in spots that are twice than that retrieved in the SSRR, makes this design a more suitable candidate for the observation of nonlinear processes. The fabricated DSRR (Fig.4g), realized on the same chip of the SSRR during the same fabrication run, shows a resonant-like optical transmission with a ~0.18 THz blue-shift and a lower Q-factor, if compared with the simulation ($\nu_{SIM}$~3.26 THz and $Q_{SIM}$~3.1). Specifically, $\nu_{EXP}$~3.44±0.4 THz and $Q_{EXP}$~2.02±0.45, exhibiting an overall behavior similar to that measured on the SSRR. Similar results were found for the optical transmission measured on the sample H1.

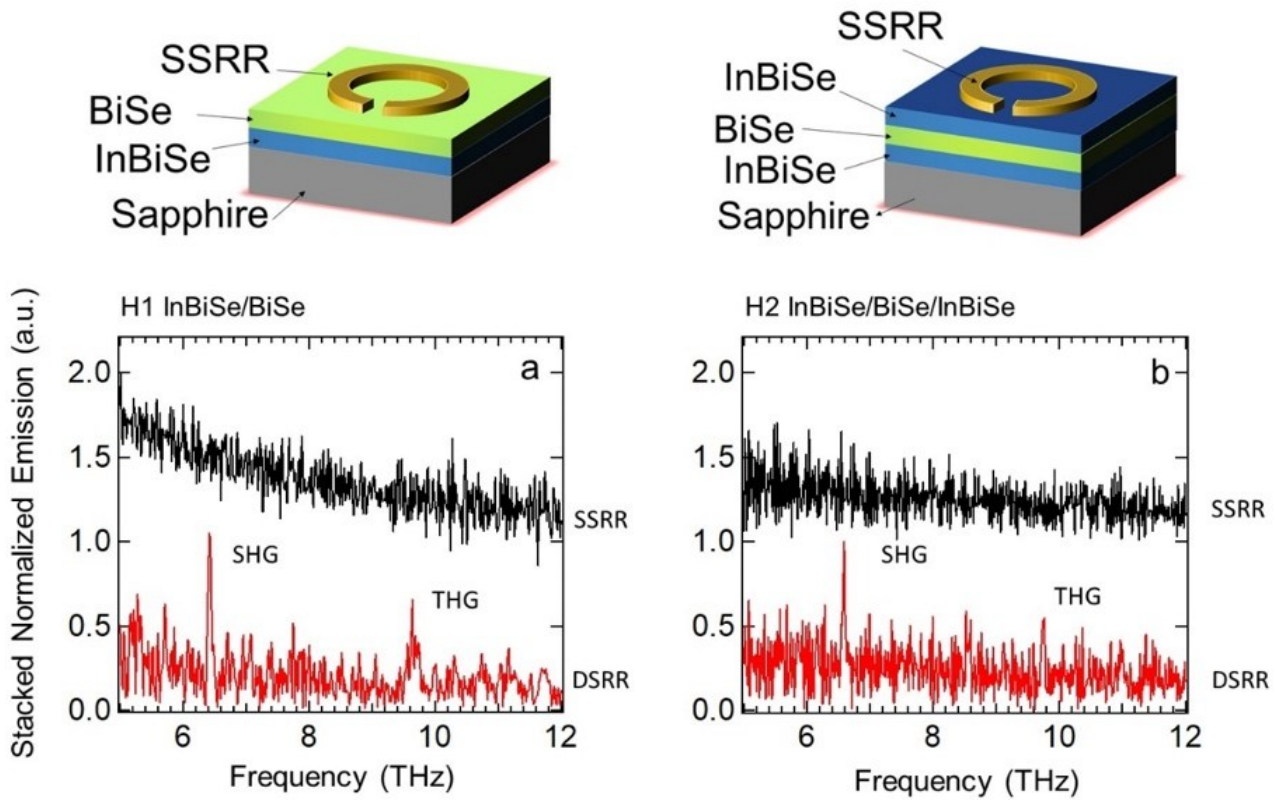

**Figure 5. (a-b)** Schematic illustration of the single-SRR array embedding on sample H1 (a) and H2 (b) **(c-d)** Emission spectra measured from the optically pumped samples (c) H1 and (d) H2, incorporating SSRR (black) and DSRR (red) design geometries, respectively. The spectra are measured in step-scan mode with a spectral resolution ~1cm$^{-1}$ and aperture size ~5mm. The total light reflected by the samples was collected through the external input window of the FTIR spectrometer, with a silicon-bolometer connected to a lock-in amplifier for detection.

To assess the non-linear origin of any possible signal emitted by the optically pumped heterostructure SRR array, we initially measure the signal detected by a Si bolometer after filtering the QCL emitted frequency with a 4 mm thick Ta-filter (see SI, section S10). To investigate possible SHG or THG in the $Bi_2Se_3$-heterostructure SRR arrays, we used the external-pump apparatus of Fig.3a, but slightly modified to collect the beam reflected by the TI surface, avoiding higher harmonic absorption by the sapphire substrate. This was achieved by mounting the TI film on a rotating stage set at a 45° angle with respect to both the incoming beam angle, and to the collection axis of the FTIR spectrometer. Under these conditions, the beam emitted by the QCL pump is focused onto the sample surface, while the reflected total signal, comprising the linear reflection and the nonlinear emission terms, both generated from the QCL pump, are collected at a 90° angle with respect to the incidence beam, and directly fed onto the FTIR spectrometer. The emission spectra are then retrieved by measuring the total signal collected from the illuminated $Bi_2Se_3$ -SRRs, after filtering out the fundamental mode.

The emission spectra are reported in Fig.5c-d, for sample H1 (Fig.5c) and H2 (Fig.5d), respectively, as acquired for the SSRR (black) and DSRR (red) resonators. Remarkably, we retrieved clear SHG peaks and slightly weaker THG peaks in both H1 and H2 for the DSRR array, i.e. the higher field enhancement design, meaning that, independently from the nature of the topological surface states (trivial or topologically protected) SHG is activated but with a stronger bulk contribution causing THG, in the sample with no topological protection (H1). No evidence of either SHG or THG was seen in the SSRRs, likely due to the fact that the field enhancement is not enough to allow a conversion efficiency in the noise limits of our set-up.

## 3. Discussion

The conversion efficiencies, determined from the experimental data, are listed in Table 1, together with the THG efficiency extracted from the previously discussed $Bi_2Se_3$ SSRR. The THG retrieved on both the H1 and H2 DSRRs are in broad agreement with the CE values found on the simple $Bi_2Se_3$ SRR transferred on $SiO_2$/Si, testifying to the very similar origin of the nonlinear process. The presence of a THG that is less intense then SHG, indicates a lower 3$^{rd}$ order conversion efficiency. This is in apparent contrast with other reports[39], and with the general property of stronger odd nonlinearities in centrosymmetric crystals.[44] Our experimental findings suggest the existence of alternative mechanisms supporting the SHG, such as IR-active phonon assisted modes[67] superimposed to the resonator-assisted complex optical response, that may increase even-order nonlinearities. However, it should be noted that the de-encoding of each single harmonic in the interferometric trace, and the procedures used to extract the spectra using Fourier transforms in the frequency domain, is not trivial, and could be masked by significant noise.

| Sample | SHG | THG |
|---|---|---|
| H1, DSRR | $(3.5 \pm 1) \times 10^{-5}$ | $(2.7 \pm 0.7) \times 10^{-5}$ |
| H2, DSRR | $(2.5 \pm 1) \times 10^{-5}$ | $(1.9 \pm 0.7) \times 10^{-5}$ |
| A, SSRR | Not detected | $(1.3 \pm 0.3) \times 10^{-5}$ |

**Table 1**. Experimental conversion efficiencies

.

| Sample | $\chi^3$ (m$^2$V$^{-2}$) |
|---|---|
| H1 | $9.2 \times 10^{-9}$ |
| H2 | $6.9 \times 10^{-9}$ |
| A | $3.31 \times 10^{-9}$ |

**Table 2**. Third order non-linear susceptibility

From the measured experimental conversion efficiencies we have extracted the third order non-linear susceptibility $\chi^3$, as follows. The conversion from the pump frequency to the TH is described by the relation:[14] $E_{TH} = \gamma L E_0^3$, where E$_0$ ~ 8.7×10$^4$ V/m is the pump field at the fundamental frequency, E$_{TH}$ is the up-converted field, L=2nm is the TI effective thickness, and $\gamma$ is related to $\chi^3$ by the equation:

$$\chi^3 = \frac{\gamma}{3/4(\pi \nu_{TH}/c n_{TH})}$$

Where $c$ is the speed of light, $\nu_{TH}$ = 9.63 THz is the TH frequency, and n$_{TH}$~1.8 is the TI refractive index at the TH. By estimating the electric field at the TH from the experimental CE, we then obtain the $\chi^3$ shown in the following table.

The extrapolated $\chi^3$ values, in the combined TI/resonator system, are of the same order of magnitude than those retrieved in graphene $\chi^{(3)}$~10$^{-9}$ m$^2$/V$^2$ , i.e. almost 7 orders of magnitude larger than THz QCL semiconductor heterostructures ($\chi^{(3)}$~7×10$^{-16}$ m$^2$/V$^2$) and more that 10 orders of magnitude higher than typical 'THz' materials.[7]

In conclusion, we designed a Bi$_2$Se$_3$-embedded SRR to drive significant field amplification in the split-gap region of a circularly shaped-ring, using resonances tuned to match the frequency of the external pump laser. THG at 9.75 THz from a ~20 thick Bi$_2$Se$_3$ layer was retrieved, at a frequency corresponding to the third order up-conversion of the fundamental laser mode. This arises from the film's nonlinear response, when dominated by the centrosymmetric properties of the Bi$_2$Se$_3$ crystal. SHG and THG at 6.5 THz and 9.75 THz, respectively, are revealed by embedding vdW (In$_x$Bi$_{(1-}$ $_x$)$_2$Se$_3$/Bi$_2$Se$_3$-heterostructures into DSSRs, which induce a two order-of-magnitude field enhancement factor. The improved film quality in the heterostructure TI led to a strong nonlinear response, mainly provided by the topological surface states, which induce crystal symmetry breaking and unlock even-order nonlinear processes. Our experiments demonstrate a novel pathway for light generation in a frequency range between the MIR and THz regions, i.e. 25 μm – 60 μm (12 – 5 THz), traditionally difficult to be exploited with practical solid-state laser technologies owing to parasitic optical phonon absorption (the *Restrahlenband*) in the constituent III-V materials used to fabricate the only miniaturized optical source adopted in this range, the QCLs. Integrating on-chip the explored novel material systems with monolithic QCL sources can be a very promising route to maximize the conversion efficiency of both THG and SHG at high (> 6 THz) THz frequencies, exploiting the huge intracavity QCL power and possible plasmonic structures lithographically devised on chip.

**Materials and methods**

**Bi$_2$Se$_3$ heterostructure TI growth and characterization**

The samples were grown on [0001] sapphire substrates. Following ref.[64], the initial (In$_x$Bi$_{(1-x)}$)$_2$Se$_3$ layer was nucleated by depositing 5 nm of Bi$_2$Se$_3$ at 235°C (temperatures measured by a thermocouple attached to the sample manipulator), followed by 5 nm of In$_2$Se$_3$ , and then annealing the stack to 325C under a Se flux to create a 10 nm thick layer (In$_x$Bi$_{(1-x)}$)$_2$Se$_3$ . After annealing, sample H1 was cooled under a Se flux and an additional 20 nm of (In$_x$Bi$_{(1-x)}$)$_2$Se$_3$ was deposited, for a 30 nm thick total layer, followed by a single, 20 nm thick Bi$_2$Se$_3$ layer. For sample H2, a similar Bi$_2$Se$_3$ layer was grown, then capped with 20 nm (In$_x$Bi$_{(1-x)}$)$_2$Se$_3$. Both samples were then diced and patterned into standard hall bars by chemical wet etching, and Cr/Au ohmic contacts were subsequently deposited before being loaded into a continuous flow helium cryostat with a base temperature of 1.6 K and an 8 T superconducting magnet. Transverse and Hall conductivities were measured using standard lock-in amplifier techniques at a frequency of 199.77 Hz.

**Nanofabrication of the Bi$_2$Se$_3$ resonator**

The Bi$_2$Se$_3$ resonators were fabricated on an a high ($10^4$ Ωcm)-resistivity 300-mm-thick Si substrate, coated with a 300 nm-thick SiO$_2$ layer (by Siltronyx). The CSRR array pattern is defined by optical lithography using a LOR3A/S1805 bilayer photoresist, on a 5×5 mm$^2$ area, followed by metal evaporation and liftoff of 10nm/100nm of Cr/Au. The 1×1 cm$^2$ BiSe film is then transferred onto the top of the SSRR array, following a PMMA assisted method[53], using a 30% KOH solution at ~70 °C to promote the delamination of the MBE grown TI from the Sapphire substrate. The Bi$_2$Se$_3$ in the gap area is then patterned with a second lithography step, followed by a plasma-Ar etching to remove the TI film from the area surrounding the gap, with a final cleaning by acetone soaking.
The resonators embedding the vdW TI heterostructures are fabricated by electron beam pattering directly on the as-grown MBE samples, employing the undercut-profile PMMA 950K polymers as a liftoff e-beam resist.

**Numerical simulations**

The SSRR single unit element (Fig.1b), comprises a perfect electric conductor (PEC) C-shaped ribbon (external radius, 4.1 μm, internal radius 3.2 μm, gap 1.5 μm) at the interface between a 300-nm-SiO$_2$/Si bottom and air top domains. The unit cell pitch is $p$ = 14.5 μm, and the total height of the calculation domain is 85 μm, i.e. 35 μm-thick dielectric substrate and 50 μm-thick volume of air. The FEM calculation (by COMSOL Multiphysics) were carried out by setting Floquet boundary conditions along the x,y side edges, and scattering boundary conditions on the top and bottom domain edges (z edges). The transmittance is calculated as $T = |S_{21}|^2$, where the S-parameters are extracted by placing one input port on top of the air domain and one receiver port on the bottom of the Si domain. A rectangular 5 × 5 μm² area around the split gap is defined, and a transition boundary condition is set based on the $Bi_2Se_3$ complex optical conductivity, calculated from the procedure described in the text, and reported in the SI. $A_G$ is calculated as the ratio between the electric field magnitude, averaged over the gap region surface, and the average field magnitude computed in absence of the resonator, from the same simulation module, eliminating the metal split ring. In the $Bi_2Se_3$-heterostructure SSRR and DSSR, the $Bi_2Se_3$ film, modeled with the same complex optical constant, was set as transition boundary layer over the entire region surrounding the metal ring.

**Optical Set-up**

The laser used in this work is a QCL emitting at 3.25 THz (λ = 108 μm), based on a bound to continuum-optical phonon hybrid active region design[68]. The 25-μm-thick GaAa/AlGaAs active region is patterned into a surface-plasmon waveguide onto a GaAs substrate, using a combination of wet etching and metal deposition. A 5-nm-thick, 40-μm-wide Ni side-absorber, overlapping on 3 μm on each side of the upper 150 nm-thick Au over-layer, is introduced on each edge of the ridge to increase the difference in losses between fundamental and higher order transverse modes in order to suppress fully the higher-order competing modes. The 700-nm-thick heavily Si-doped ($5 \times 10^{18}$ cm$^{-3}$) GaAs top contact layer lies between the active region and the substrate.

The QCL, driven in pulsed-mode (1 μs pulse, 50 kHz repetition rate) and emitting 2.5 W peak power is focused onto a ~0.4 mm spot on the surface of the $Bi_2Se_3$-resonators array facing the external-input window of an FTIR spectrometer (Bruker Vertex V80). The THG or SHG signal is isolated after filtering the QCL emission with a 6 THz high bandpass 2 mm-thick Ta-filter. The spectra are acquired under vacuum, in step scan mode, with a resolution of 1 cm$^{-1}$. The signal is measured by a helium-cooled Si-bolometer, connected to a lock-in amplifier, referenced to an amplitude modulated signal at 317 Hz, that is also simultaneously gating the QCL bias pulses.

**Acknowledgements.** We acknowledge funding from the European Union through the FET Open project EXTREME IR (944735), and the UK Research and Innovation councils through the grant 'NAME' (EP/V001914/1). The samples were grown in the Royce Deposition system at the University of Leeds, which was funded by the Henry Royce Institute, UK, through grant EP/P022464/1.

**Conflict of interests**

The authors declare no competing financial and non-financial interests.

**Data availability:** The data presented in this study are available on reasonable request from the corresponding author.

**Code availability statement:** The relevant computer codes supporting this study are available from the authors upon reasonable request.

**Contributions**

M.S.V. and A.D. conceived the concept and designed the experiment; A.D. fabricated the samples, performed the experiments and analyzed the data. S.G. performed the simulations. C.K., A. Y., S. S. and J.F. grew the TI heterostructures and performed magneto-spectroscopy experiments. M.S., L.L., E. H. L, grew the QCL heterostructure. A.D. and M.S.V. wrote the manuscript. M.S.V. supervised the study. All authors analysed the data, discussed the results and contributed to the writing of the manuscript.

# References

1. Geim, A. K. & Novoselov, K. S. The rise of graphene. *Nat. Mater.* **6**, 183–191 (2007).
2. Hasan, M. Z. & Moore, J. E. Three-Dimensional Topological Insulators. *Annu. Rev. Condens. Matter Phys.* **2**, 55–78 (2011).
3. Moore, J. E. The birth of topological insulators. *Nature* **464**, 194–198 (2010).
4. R., N. R. *et al.* Fine Structure Constant Defines Visual Transparency of Graphene. *Science (80-. ).* **320**, 1308 (2008).
5. Tomadin, A. *et al.* The ultrafast dynamics and conductivity of photoexcited graphene at different Fermi energies. *Sci. Adv.* **4**, eaar5313 (2023).
6. Brida, D. *et al.* Ultrafast collinear scattering and carrier multiplication in graphene. *Nat. Commun.* **4**, 1987 (2013).
7. Hafez, H. A. *et al.* Terahertz Nonlinear Optics of Graphene: From Saturable Absorption to High-Harmonics Generation. *Adv. Opt. Mater.* **8**, 1900771 (2020).
8. Bianchi, V. *et al.* Terahertz saturable absorbers from liquid phase exfoliation of graphite. *Nat. Commun.* **8**, 1–9 (2017).
9. Marini, A., Cox, J. D. & García de Abajo, F. J. Theory of graphene saturable absorption. *Phys. Rev. B* **95**, 125408 (2017).
10. Vermeulen, N. *et al.* Graphene's nonlinear-optical physics revealed through exponentially growing self-phase modulation. *Nat. Commun.* **9**, 2675 (2018).
11. Gu, T. *et al.* Regenerative oscillation and four-wave mixing in graphene optoelectronics. *Nat. Photonics* **6**, 554–559 (2012).
12. Castro Neto, A. H., Guinea, F., Peres, N. M. R., Novoselov, K. S. & Geim, A. K. The electronic properties of graphene. *Rev. Mod. Phys.* **81**, 109–162 (2009).
13. Soavi, G. *et al.* Broadband, electrically tunable third-harmonic generation in graphene. *Nat. Nanotechnol.* **13**, 583–588 (2018).
14. Hafez, H. A. *et al.* Extremely efficient terahertz high-harmonic generation in graphene by hot Dirac fermions. *Nature* **561**, 507–511 (2018).
15. Yoshikawa, N., Tamaya, T. & Tanaka, K. High-harmonic generation in graphene enhanced by elliptically polarized light excitation. *Science (80-. ).* **356**, 736–738 (2017).
16. Mao, W., Rubio, A. & Sato, S. A. Enhancement of high-order harmonic generation in graphene by mid-infrared and terahertz fields. *Phys. Rev. B* **109**, 45421 (2024).
17. Soavi, G. *et al.* Hot Electrons Modulation of Third-Harmonic Generation in Graphene. *ACS Photonics* **6**, 2841–2849 (2019).
18. Cox, J. D., Marini, A. & de Abajo, F. J. G. Plasmon-assisted high-harmonic generation in graphene. *Nat. Commun.* **8**, 14380 (2017).
19. Sirtori, C., Capasso, F., Sivco, D. L. & Cho, A. Y. Giant, triply resonant, third-order nonlinear susceptibility ${\mathrm{\ensuremath{\chi}}}_{3\mathrm{\ensuremath{\omega}}}^{(3)}$ in coupled quantum wells. *Phys. Rev. Lett.* **68**, 1010–1013 (1992).
20. Vermeulen, N. *et al.* Post-2000 nonlinear optical materials and measurements: data tables and best

practices. *J. Phys. Photonics* **5**, 35001 (2023).

21. Hafez, H. A. *et al.* Intense terahertz field effects on photoexcited carrier dynamics in gated graphene. *Appl. Phys. Lett.* **107**, 251903 (2015).

22. Chiadroni, E. *et al.* The SPARC linear accelerator based terahertz source. *Appl. Phys. Lett.* **102**, 94101 (2013).

23. Abedinpour, S. H. *et al.* Drude weight, plasmon dispersion, and ac conductivity in doped graphene sheets. *Phys. Rev. B* **84**, 45429 (2011).

24. Frenzel, A. J., Lui, C. H., Shin, Y. C., Kong, J. & Gedik, N. Semiconducting-to-Metallic Photoconductivity Crossover and Temperature-Dependent Drude Weight in Graphene. *Phys. Rev. Lett.* **113**, 56602 (2014).

25. Low, T. & Avouris, P. Graphene Plasmonics for Terahertz to Mid-Infrared Applications. *ACS Nano* **8**, 1086–1101 (2014).

26. Messelot, S., Coeymans, S., Tignon, J., Dhillon, S. & Mangeney, J. High Q and sub-wavelength THz electric field confinement in ultrastrongly coupled THz resonators. *Photon. Res.* **11**, 1203–1216 (2023).

27. Di Gaspare, A. *et al.* Compact terahertz harmonic generation in the Reststrahlenband using a graphene-embedded metallic split ring resonator array. *Nat. Commun.* **15**, 2312 (2024).

28. Qi, X.-L. & Zhang, S.-C. Topological insulators and superconductors. *Rev. Mod. Phys.* **83**, 1057–1110 (2011).

29. Hoefer, K. *et al.* Intrinsic conduction through topological surface states of insulating Bi2Te3 epitaxial thin films. *Proc. Natl. Acad. Sci.* **111**, 14979–14984 (2014).

30. Pistore, V. *et al.* Terahertz Plasmon Polaritons in Large Area Bi2Se3 Topological Insulators. *Adv. Opt. Mater.* **n/a**, 2301673 (2023).

31. Pogna, E. A. A. *et al.* Mapping propagation of collective modes in Bi2Se3 and Bi2Te2.2Se0.8 topological insulators by near-field terahertz nanoscopy. *Nat. Commun.* **12**, 6672 (2021).

32. Chen, S. *et al.* Real-space nanoimaging of THz polaritons in the topological insulator Bi2Se3. *Nat. Commun.* **13**, 1374 (2022).

33. Low, T. *et al.* Polaritons in layered two-dimensional materials. *Nat. Mater.* **16**, 182–194 (2017).

34. Basov, D. N., Fogler, M. M. & García de Abajo, F. J. Polaritons in van der Waals materials. *Science (80-. ).* **354**, aag1992 (2016).

35. He, J. & Tritt, T. M. Advances in thermoelectric materials research: Looking back and moving forward. *Science (80-. ).* **357**, eaak9997 (2017).

36. Cui, B. *et al.* Low-Power and Field-Free Perpendicular Magnetic Memory Driven by Topological Insulators. *Adv. Mater.* **35**, 2302350 (2023).

37. Wang, Y. *et al.* Enhancement of spintronic terahertz emission enabled by increasing Hall angle and interfacial skew scattering. *Commun. Phys.* **6**, 280 (2023).

38. Liu, X. & Hersam, M. C. 2D materials for quantum information science. *Nat. Rev. Mater.* **4**, 669–684 (2019).

39. Stensberg, J. *et al.* Observation of terahertz second harmonic generation from Dirac surface states in the topological insulator ${\mathrm{Bi}}_{2}{\mathrm{Se}}_{3}$. *Phys. Rev. B* **109**, 245112 (2024).

40. Tielrooij, K.-J. *et al.* Milliwatt terahertz harmonic generation from topological insulator metamaterials. *Light Sci. Appl.* **11**, 315 (2022).

41. McIver, J. W. *et al.* Theoretical and experimental study of second harmonic generation from the surface of the topological insulator Bi 2Se 3. *Phys. Rev. B - Condens. Matter Mater. Phys.* **86**, 1–11 (2012).

42. Stensberg, J. *et al.* Observation of terahertz second harmonic generation from Dirac surface states in the topological insulator Bi$_2$Se$_3$. (2023).

43. Niesler, F. B. P., Feth, N., Linden, S. & Wegener, M. Second-harmonic optical spectroscopy on split-ring-resonator arrays. *Opt. Lett.* **36**, 1533–1535 (2011).

44. Bai, Y. *et al.* High-harmonic generation from topological surface states. *Nat. Phys.* **17**, 311–315 (2021).

45. Chen, X., Ma, X.-C., He, K., Jia, J.-F. & Xue, Q.-K. Molecular Beam Epitaxial Growth of Topological Insulators. *Adv. Mater.* **23**, 1162–1165 (2011).

46. Menshchikova, T. V *et al.* Band Structure Engineering in Topological Insulator Based Heterostructures. *Nano Lett.* **13**, 6064–6069 (2013).

47. Pistore, V. *et al.* Holographic Nano-Imaging of Terahertz Dirac Plasmon Polaritons Topological Insulator Antenna Resonators. *Small* 2308116 (2023) doi:10.1002/SMLL.202308116.

48. Li, Y., Bowers, J. W., Hlevyack, J. A., Lin, M.-K. & Chiang, T.-C. Emergent and Tunable Topological Surface States in Complementary Sb/Bi(2)Te(3) and Bi(2)Te(3)/Sb Thin-Film Heterostructures. *ACS Nano* **16**, 9953–9959 (2022).

49. Padilla, W. J., Basov, D. N. & Smith, D. R. Negative refractive index metamaterials. *Mater. Today* **9**, 28–35 (2006).

50. Al-Naib, I. & Withayachumnankul, W. Recent Progress in Terahertz Metasurfaces. *J. Infrared, Millimeter, Terahertz Waves* **38**, 1067–1084 (2017).

51. Kim, S. *et al.* High-harmonic generation by resonant plasmon field enhancement. *Nature* **453**, 757–760 (2008).

52. You, J. W. & Panoiu, N. C. Polarization control using passive and active crossed graphene gratings. *Opt. Express* **26**, 1882–1894 (2018).

53. Bansal, N. *et al.* Transferring MBE-Grown Topological Insulator Films to Arbitrary Substrates and Metal–Insulator Transition via Dirac Gap. *Nano Lett.* **14**, 1343–1348 (2014).

54. Massicotte, M., Soavi, G., Principi, A. & Tielrooij, K.-J. Hot carriers in graphene – fundamentals and applications. *Nanoscale* **13**, 8376–8411 (2021).

55. Han, J. W. *et al.* Plasmonic Terahertz Nonlinearity in Graphene Disks. *Adv. Photonics Res.* **3**, 2100218 (2022).

56. Jadidi, M. M. *et al.* Nonlinear Terahertz Absorption of Graphene Plasmons. *Nano Lett.* **16**, 2734–2738 (2016).

57. Di Gaspare, A. *et al.* Electrically Tunable Nonlinearity at 3.2 Terahertz in Single-Layer Graphene. *ACS Photonics* (2023) doi:10.1021/acsphotonics.3c00543.

58. Sobota, J. A. *et al.* Ultrafast electron dynamics in the topological insulator Bi2Se3 studied by time-resolved photoemission spectroscopy. *J. Electron Spectros. Relat. Phenomena* **195**, 249–257 (2014).

59. Mics, Z. *et al.* Thermodynamic picture of ultrafast charge transport in graphene. *Nat. Commun.* **6**, 7655 (2015).

60. Zhang, H. *et al.* Ultrafast saturable absorption in topological insulator Bi2SeTe2 nanosheets. *Opt. Express* **23**, 13376–13383 (2015).

61. Yan, P. *et al.* A practical topological insulator saturable absorber for mode-locked fiber laser. *Sci.*

*Rep.* **5**, 8690 (2015).

62. Guo, T., Jin, B. & Argyropoulos, C. Hybrid Graphene-Plasmonic Gratings to Achieve Enhanced Nonlinear Effects at Terahertz Frequencies. *Phys. Rev. Appl.* **11**, 24050 (2019).

63. Jadidi, M. M. *et al.* Tunable Terahertz Hybrid Metal–Graphene Plasmons. *Nano Lett.* **15**, 7099–7104 (2015).

64. Wang, Z. & Law, S. Optimization of the Growth of the Van der Waals Materials Bi2Se3 and (Bi0.5In0.5)2Se3 by Molecular Beam Epitaxy. *Cryst. Growth Des.* **21**, 6752–6765 (2021).

65. Spirito, D. *et al.* Weak antilocalization and spin-orbit interaction in a two-dimensional electron gas. *Phys. Rev. B* **85**, 235314 (2012).

66. Liu, P. Q. *et al.* Highly tunable hybrid metamaterials employing split-ring resonators strongly coupled to graphene surface plasmons. *Nat. Commun.* **6**, 1–7 (2015).

67. Richter, W., Kohler, H. & Becker, C. A Raman and Far-Infrared Investigation of Phonons. *Phys. Stat. sol* **84**, 619 (1977).

68. Song, C. *et al.* High-power density, single plasmon, terahertz quantum cascade lasers via transverse mode control. *Appl. Phys. Lett.* **122**, 121108 (2023).

# Supporting Information

# Second harmonic generation and third harmonic generation in topological insulator-based van der Waals metamaterials

**Alessandra Di Gaspare[1], Sara Ghayeb,[1] Craig Knox[2] Edmund H. Linfield[2], Joshua Freeman,[2] and Miriam S. Vitiello[1]**

[1]*NEST, CNR-NANO and Scuola Normale Superiore, 56127, Pisa, Italy*
[2]*School of Electronic and Electrical Engineering, University of Leeds, Leeds, LS2 9JT, UK*

**S1. Bi$_2$Se$_3$ Optical Constant**

The ~20nm thick Bi$_2$Se$_3$ film of Section 2-3 of the main article has been characterized by low-temperature magneto transport[1]. The fitting analysis of the experimental magnetoresistance curves reveal the presence of two surface carrier bands, the first having lower mobility and density of carriers, attributed to the trivial surface stated (2D massive electrons) and the second associated with the topological surface states. Each one of these two band can be treated independently, providing its own contribution to the optical conductivity, namely:

$$\sigma_{DM}(\nu) = \frac{-iD_0}{\pi} \frac{1}{(2\pi\nu + i\Gamma_0)} \tag{S1}$$

$\sigma_{DM}(\nu)$ is here the intraband, Drude-like complex conductivity of the Bi$_2$Se$_3$ topological Dirac material (DM) forming the surface state, where $D_0 = E_F e^2/\hbar^2$ is the linear Drude weight, E$_F$ the Fermi energy, $e$ is the electron charge, $\hbar$ is the reduced Planck constant, $\Gamma_0 = \tau_0^{-1} = ev_F^2/E_F\mu$ is the scattering rate, v$_F$ the Fermi velocity, and μ is the carrier mobility. For the Bi$_2$Se$_3$ topological state, we set v$_F$ = 5×10$^5$ m/s, $E_F = \hbar v_F \sqrt{4\pi n_{DM}}$, with $n_{DM}$ the carrier density of the topological surface state. From the magnetotransport measurements, we extract the values μ=1000 cm$^2$/Vs, and $n_{DM}$=0.65×10$^{13}$ cm$^{-2}$, and we obtain E$_F$ = 296 meV.

$$\sigma_{2DEG}(\nu) = \frac{i}{2\pi\nu} \frac{n_{2DEG} e^2}{m^*} \tag{S2}$$

$\sigma_{2DEG}(\nu)$ is the complex conductivity of the 2D classical electron gas, where m*=0.15m$_e$ in Bi$_2$Se$_3$ trivial surface state, and n$_{2DEG}$ = 1.9×10$^{13}$ cm$^{-2}$ is 2D carrier density extracted from the massive electron band contribution to the Hall magnetoresistance. The surface state contribute to the total permittivity is then:

$$\varepsilon_{2D,tot} = 1 + 2 \times \left(\frac{\sigma_{DM} + \sigma_{2DEG}}{2\pi\nu\varepsilon_0 t_{TI}}\right) \tag{S3}$$

where the factor 2 accounts for the topological surfaces on both sides of the Bi$_2$Se$_3$ slab, $\varepsilon_0$=8.85×10$^{-12}$ F/m is the vacuum permittivity, and t$_{TI}$ ~ 1nm is the quantum layer thickness of the Bi$_2$Se$_3$ surface state.

The 20 nm thick bulk has a permittivity given by the sum of the Drude free carrier primitivity, $\varepsilon_{Drude}$, and three Lorenz-Drude terms arising from the E$_u$ and A$_{2u}$ TO phonons, and the bandgap absorption oscillator[1,2]:

$$\varepsilon_{Drude} = \frac{\omega_{Drude}^2}{\omega^2+i\omega\gamma_{Drude}} \quad (S4)$$

$$\varepsilon_{Bulk} = \varepsilon_{Drude} + \sum_{j=1}^{3}\frac{\omega_{pj}^2}{\omega_{0j}^2-\omega^2-i\omega\gamma_j} \quad (S5)$$

In table 1, all the parameters appearing in eqs.S4-S5 are listed.

| Parameter | Value | Description |
| --- | --- | --- |
| $\omega_{Drude}$ | 900 cm$^{-1}$ | Free carrier plasma frequency BiSe bulk |
| $\gamma_{Drude}$ | 733 cm$^{-1}$ | Drude damping rate |
| $\omega_{p1}$ | 675.9 cm$^{-1}$ | TO phonon (E$_u$) oscillator strength |
| $\omega_{01}$ | 63.03 cm$^{-1}$ | TO phonon frequency |
| $\gamma_{01}$ | 17.05 cm$^{-1}$ | TO phonon damping rate |
| $\omega_{p2}$ | 100 cm$^{-1}$ | TO phonon (A$_{2u}$) oscillator strength |
| $\omega_{02}$ | 127 cm$^{-1}$ | TO phonon frequency |
| $\gamma_{02}$ | 10 cm$^{-1}$ | TO phonon damping rate |
| $\omega_{p3}$ | 11249 cm$^{-1}$ | bandgap oscillator strength |
| $\omega_{03}$ | 2029.5 cm$^{-1}$ | bandgap frequency |
| $\gamma_{03}$ | 117.3 cm$^{-1}$ | bandgap damping rate |

**Table 1.** Optical parameters of BiSe[1,2]

The Bi$_2$Se$_3$ total permittivity is then $\varepsilon_{BiSe} = \varepsilon_{Bulk} + \varepsilon_{2D}$.

The complex refractive index (n,k) is then extracted from the relationships:

$$n = Re\left(\sqrt{\varepsilon_{BiSe}}\right); k = Im\left(\sqrt{\varepsilon_{BiSe}}\right) \quad (S6)$$

The real and imaginary parts of the complex refractive index are reported in fig. S1a. In Fig.S1b, is shown only the contribution of the bulk, that will be used in the third harmonic generation numerical model, to account for the bise nonlinear response from the surface states, with the Bi$_2$Se$_3$ bulk providing the dielectric environment for the nonlinear, surface current generator.

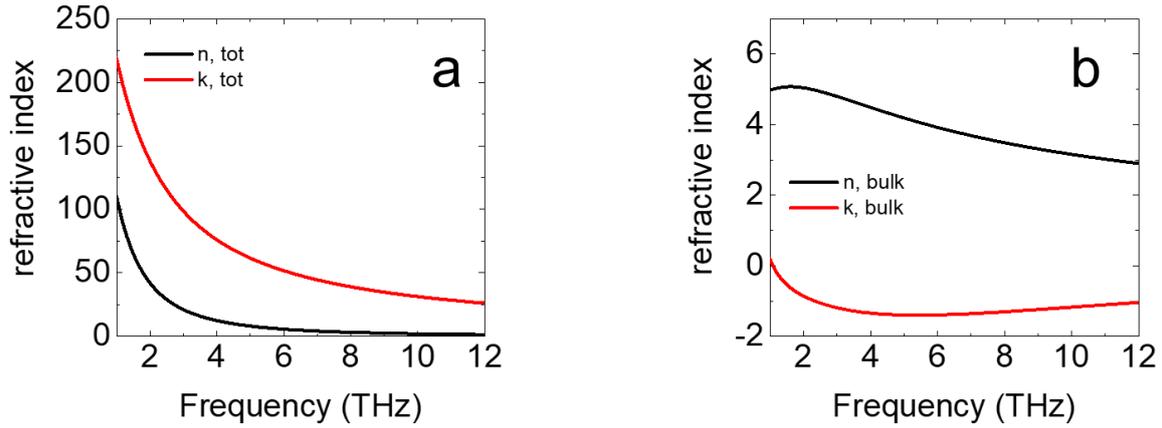

**Figure S1.** a) Complex refractive index of the $Bi_2Se_3$ film, calculated from the total permittivity, comprising the surface states and the bulk contributions. b) Complex refractive index of the $Bi_2Se_3$ bulk.

**S2. Effect of the TI film bending**

The large area Bi2Se3 film is robust after the wet chemical detaching, and upon transfer the damage or bending is usually negligible. However, to evaluate any possible influence on the optical properties that could affect the resonator response, in Fig.S2 (top) we compare the simulations performed by considering Bi2Se3 as suspended and with a slightly bent surface across the two metal edges of the resonator gap, modeled as 100-nm thick Au (optical properties defined in COMSOL library), with the case of Bi2Se3 film in contact with the substrate below (below), not showing any significant difference.

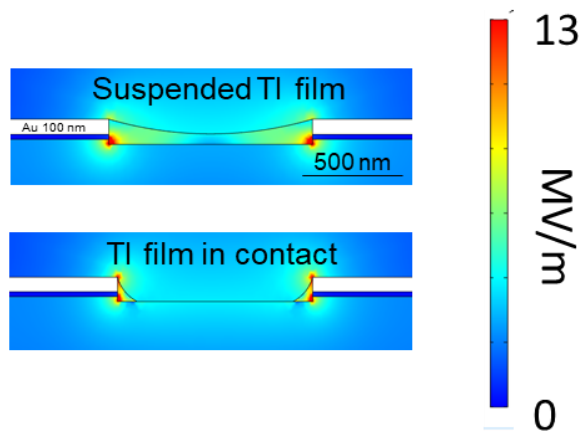

**Figure S2:** Electric field distribution across the gap area of the $Bi_2Se_3$-SSRR, calculated using FEM simulations in the two cases: suspended film (top), and in contact with the substrate (bottom)

## S3. Single split ring resonator: field amplification

The response of the single split ring resonator (SSRR) of Section 2-3, when illuminated by a plane wave linearly polarized along the direction indicated by the arrow in Fig.1b, leads to the formation of a hot spot, i.e. a region where the e.m. field intensity is up to two orders of magnitude higher than the field of the incoming beam. The resonator essentially provides a field enhancement, at the resonance, in the split gap area, owing to the dipolar mode established in the region enclosed by the two metal arms. The resonance frequency and the field enhancement can be tuned, by design, through the ring geometry; however, they are also affected by the dielectric response of the split gap region. To account for the presence of the thin $Bi_2Se_3$ layer, and to highlight its role in determining the field amplification needed to enhance the frequency up-conversion efficiency, in Fig.S2 we compare the gap amplification simulated in the bare SSRR (black) and in the SSRR with $Bi_2Se_3$ in the gap (red). The bare SSRR was realized by the metal ring, modeled as a perfect electric conductor (PEC), on top of a 300nm-$SiO_2$/Si substrate, shaped as a THz-transparent dielectric material ($n_{SiO2}$~2, $n_{Si}$ ~3.4). The $Bi_2Se_3$ integrated SSRR comprises a 5×5μm$^2$ area around the gap, set as a transition boundary condition with the complex refractive index introduced in Section S1.

The gap amplification peak is slightly blue-shifted and decreased in the BiSe/SSRR, as a consequence of the different optical contrast between the materials of the resonators, particularly localized in the split gap, i.e. the most sensitive area to electromagnetic changes.

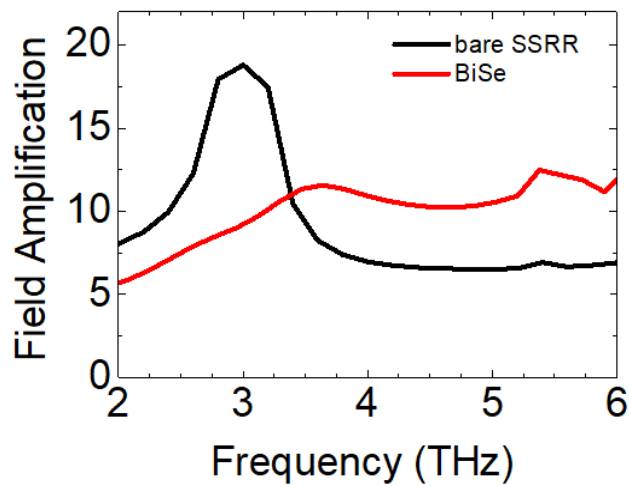

**Figure S3**. Comparison of the field amplification gap simulated on the bare SSRR (black) and in the $Bi_2Se_3$/SSRR, with the $Bi_2Se_3$ covering the 5×5μm$^2$ area around the gap area, described as a transition optical layer with n,k shown in Fig. S1.

## S4. Hot electron nonlinear response in $Bi_2Se_3$

In the THz range, the optical response of the $Bi_2Se_3$ is ruled by the intraband absorption[3]. The presence of a high-power optical beam drives the system to a non-equilibrium state with an excess distribution of carriers at the energy of the optical pump. This leads to the nonlinear response of the system[4], that is then dominated by the thermal electrodynamic processes resulting in the Dirac carriers, on a timescale comparable to the light oscillations, as reported for single layer graphene[5,6,7,8]. At first ultrafast (~20 fs[4]) carrier-carrier scattering determines the initial energy redistribution, bringing the system into a non-equilibrium state with electrons sharing an hot-electron temperature $T_e$.[6] Then, the system relaxes back to an equilibrium state[7,8], depending on the relaxation cooling channels available. In the present work, the pump source varies on a timescale (~μs pulse duration, 50 kHz repetition rate) much longer than any possible relaxation channel, leading the system into a steady excitation state at $T_e$ is realized[9,10], where:

$$T_e = T_{sub} + \frac{P_{in}\tau_{cool}}{C_e} \quad (S7)$$

with $T_{sub}$=300K is the temperature at equilibrium, $P_{in}$ is the excitation power intensity, in units of Wcm$^{-2}$, and $\tau_{cool}$ is the cooling time. We express the heat capacitance, $C_e$, as in the case of single layer graphene, for the limited doping regime ($E_F << K_B T_e$) as[9]:

$$C_{e,doped} = \frac{2\pi E_F}{3(\hbar v_F)^2} k_B^2 T_e \quad (S8)$$

We assume $\tau_{cool}$=3.5 ps[11], and by combining Eqs. S7 and S8, we can calculated the $T_e$ dependence on the pump power density, by using the m/s parameters of the BiSe film, $v_F$ =5×10$^5$ $E_F$ =290 meV.[1]

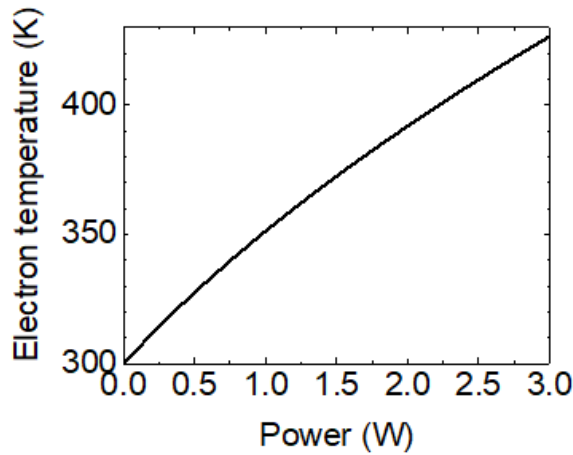

**Figure S4:** Hot-electron temperature in the $Bi_2Se_3$ pumped optically with a pump beam focused on a 0.7 mm spot, as a function of the power, according to Eqs. S6-7.

## S5. Third order field-dependent conductivity

In presence of an intense excitation beam, the SLG nonlinear response can be expressed through a field-dependent conductivity[10]:

$$\sigma_{tot}(\nu) = \sigma_0(\nu) + |E_0(\nu)|^2 \times \sigma_3(\nu) \tag{S9}$$

where $E_0$ is the field intensity, and $\sigma_0(\nu)$ is the Bi$_2$Se$_3$ linear total conductivity is calculated in Section S1. The nonlinear term of the conductivity, $\sigma_3$, is the Kerr conductivity[6]:

$$\sigma_3(\nu) = \eta[\sigma_{Kerr}(\nu)] \tag{S10}$$

whose numerical expressions are[12]:

$$\sigma_{Kerr}(\nu = \nu_{eff}) = \frac{i9e^6 v_F}{4\pi\hbar^4} \frac{D_{he}}{(2\pi\nu + i\Gamma_{he})(-2\pi\nu + i\Gamma_{he})(4\pi\nu + i\Gamma_{he})} \tag{S11}$$

where the substitution of $\nu$ with $\nu_{eff} = (\nu^2 - \nu_0^2)/\nu$ accounts for the plasmonic LC resonance of the SRR, and $\eta \sim 0.6$ is the total filling factor of SRR array. The $E_F$-dependent parameters $D_{he}$ and $\Gamma_{he}$ are the hot electron Drude weight and scattering rate, can be written as[13,12]:

$$D_{he} = D_0 \left[1 - \frac{1}{6}\left(\frac{\pi k_B}{E_F}\right)^2 T_e^2\right] \tag{S12}$$

$$\Gamma_{he} = \Gamma_0 \left[1 + \frac{1}{6}\left(\frac{\pi k_B}{E_F}\right)^2 T_e^2\right] \tag{S13}$$

The hot electron optical response is then captured by the transmittance, as for Eq.1 of the main text.

## S6. Refined numerical model for third harmonic generation (THG)

In the refined model, the THG conversion efficiency (CE) is extracted by setting up 2D simulations of the SRR array, defining the unit cell with the same geometrical configuration used for the extraction linear parameters (resonance frequency, Q-factor and field amplification) presented in Section 2 of the main article, and using frequency domain interface in (COMSOL Multiphysics Inc). A perfectly matched layer (PML) boundary condition in used in the $z$ direction defined in Fig.3d (manuscript) while periodic boundary condition was applied along the $x,y$ directions. The Bi$_2$Se$_3$ is included by defining a 20-nm thick layer with the $n,k$ shown in Fig.S1b, i.e. modeling the Bi$_2$Se$_3$ insulating bulk, with two surface current density generators on both the top Bi$_2$Se$_3$/air and the bottom Bi$_2$Se$_3$/SiO$_2$ interfaces. The current generators are first used to generate the linear electric field $E_{FH}$, relying on the linear conductivity

$$J_0 = \sigma_0 E_{FH} \tag{S14}$$

then, during a second simulation iteration, they generate the third harmonic field $E_{TH}$, relying on the field-dependent Kerr conductivity of Eq.S9, from the following equations S15-16:

$$J_{tot} = \sigma_0 E_{TH} + J_3 \qquad (S15)$$

where

$$J_3(\nu) = \sigma_3(\nu)[2|E_{FH}(\nu)|^2 E_{FH}(\nu) + c.c.]/3 \qquad (S16)$$

To calculate the CE associated to THG, we set an incident plane wave with transverse magnetic (TM) polarization and input power density $I_0$ (W/m$^2$), irradiating the resonator plane. This is corresponding to an equivalent electric field: $E = \sqrt{Z_0 I_0}$, where $Z_0 = 377\ \Omega$ is the vacuum impednacne, and $I_0$ is set in agreement with the relation: $I_0 = \frac{P_{in}}{A_{spot}}$, with $A_{spot}$ ~0.35 mm$^2$ is the illuminated spot area. The TH signal is generated by the Bi$_2$Se$_3$ embedded in the SRR, and irradiated back in the free space. CE is calculated as $P_{TH}/P_{in}$ where $P_{TH}$ is the power outflow of the TH wave, calculated by the module of the Poynting vector of the TH wave.

## S7. InBiSe optical constants

The field amplification in the split gap of the resonator is sensitive, to a given extent, to the presence of the integrated TI film, as it is relying mainly on the metallic micro-structured resonators. We first consider the optical conductivity of the InBiSe/BiSe film (In contend x=0.5), by adding a Drude-like term to the complex permittivity, accounting for the trivial 2DEG in the InBiSe and by scaling the bulk contribution following the parameters in ref.[14]

$$\varepsilon(\nu) = \varepsilon_\infty \left(1 - \frac{\nu_{Drude}^2}{\nu^2 + i\gamma_{Drude}\nu}\right) \qquad (eq.S17)$$

The list of parameters used for the calculation is shown in table 2, and the as calculated complex refractive index in the frequency range of interest is shown in fig.S5a.

| Parameter | Value | Description |
|---|---|---|
| $\nu_{Drude}$ | 2.64 THz | Free carrier plasma frequency BiSe bulk |
| $\gamma_{Drude}$ | 14 THz | Drude damping rate |
| $\varepsilon_\infty$ | 18.8 | High frequency permittivity |

**Table 2.** Optical parameters of InBiSe (from ref.[14])

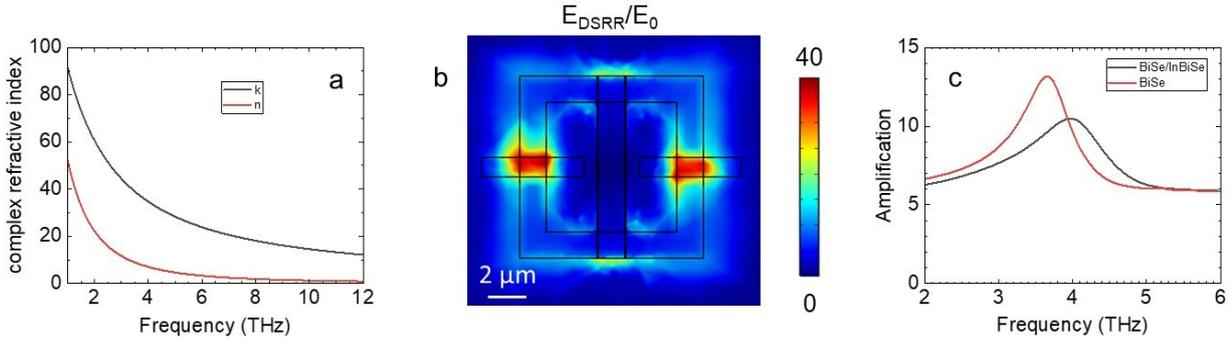

**Figure S5:** a) Complex refractive index calculated for the complete BiSe/InBiSe system (In content x=0.5). b) 2D map of the field amplification at the resonance, calculated for the DSRR c) Field amplification extracted from the simulation of the DSRR, performed considering the DSRR coated with a transistion boundary condition having the complex refractive index of the BiSe-only shown in Fig.S1a (red) and BiSe/InBiSe shown in (a) (black).

### S8. Sapphire optical constants

The response of the split ring resonators realized on the as-grown complex TI heterostructure samples of is affected by the different dielectric substrate. The higher sapphire refractive index $n_{AlO}$~3.1[15], if compared with the $n_{SiO2}$, implies a rescaling of the geometry such that the ring geometry needs to be roughly ~ $n_{AlO}/ n_{SiO2}$ smaller to match the same frequency of the pump laser. Moreover, the sapphire absorption at higher frequency must be considered, particularly in the prediction of the harmonic generation conversion efficiency. We model the absorption of the Sapphire by following the ref.[15], and generate a numerical absorption coefficient α[cm$^{-1}$], in quantitative agreement with their measured behavior. Then we set a frequency independent real part of the refractive index at $n_{AlO}$~3.1, and we extract the imaginary part from $k = \alpha c/4\pi\nu$, where c is the speed of light. The results of such model are shown in Fig.S4a, reporting the absorption, and Fig.S4b, reporting the refractive index, in the frequency range of interest in the present work.

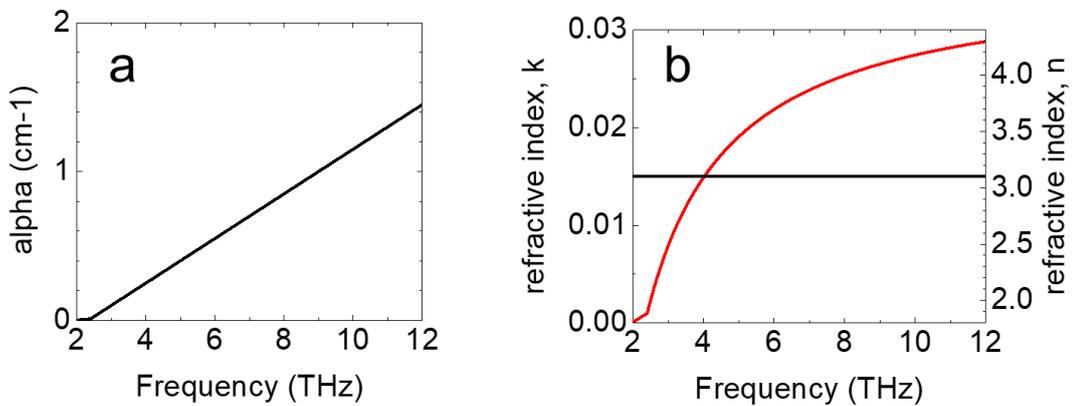

**Figure S5**: **(a)** Absorption coefficient of the Sapphire substrate in the frequency range of interest. **(b)** Complex refractive index of the Sapphire.

## S8. Micro Raman in (In$_x$Bi$_{(1-x)}$)$_2$Se$_3$/ Bi$_2$Se$_3$ heterostructures

Micro-Raman spectroscopy experiments were performed to characterize the as grown heterostructures embedded in sample H1 and H2. Raman spectra (Horiba, Explora Plus microscope) were measured using a 532 nm laser delivering 1 mW optical power, focused to a ~2μm spot. In the first case (Fig.S5a, H1), the $E_g^1$ $A_{1g}^1$, $A_{1g}^2$ and $E_g^2$ Raman modes of the Bi$_2$Se$_3$ are observed at wavenumbers of 33.1, 66.3, 170.4 and 127.0 cm$^{-1}$, respectively. For H2 (Fig.S5b), the picture is less trivial. The foremost Raman peaks of the Bi$_2$Se$_3$ are still visible, albeit over a significant, broad band background: $E_g^1$ $A_{1g}^1$, $A_{1g}^2$ and $E_g^2$ sit at 30.2, 65.9, 165.9 and 127.6 cm$^{-1}$, respectively. Three weaker peaks at 76.4, 159 and 182.3 cm$^{-1}$ are also detected, indicating the presence of weaker vibrational modes. The lower energy peaks may reflect the IR-active modes ($E_u$ and $A_{1u}$) of the Bi$_2$Se$_3$[16], or they may be replicas of the $A_{1g}^1$ and the $A_{1g}^2$ peaks, related to the (In$_x$Bi$_{(1-x)}$)$_2$Se$_3$ layer or the (In$_x$Bi$_{(1-x)}$)$_2$Se$_3$/Bi$_2$Se$_3$ heterointerface in sample H2.

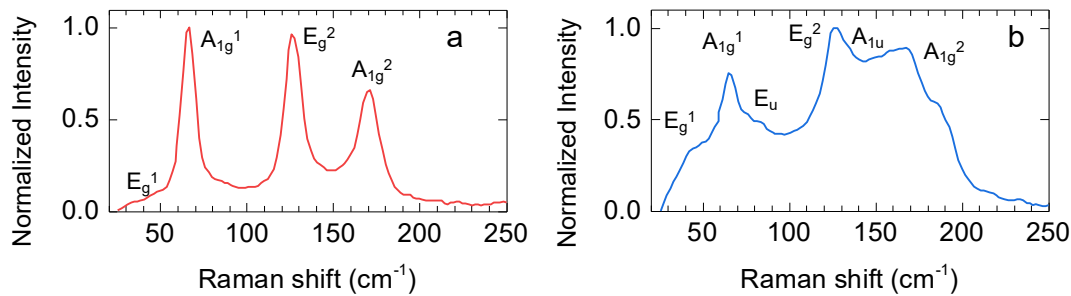

**Figure S6**: **(a-b)** Micro-Raman spectra for the as grown Bi$_2$Se$_3$-heterostructures (a) H1 and (b) H2.

## S9. Magneto-spectroscopy

To shed light on the topological nature of the heterostructure-TIs, we conducted low-temperature magneto-transport measurements. Both samples show resistances that decrease on cooling, indicating that scattering within these samples is dominated by electron-phonon interactions at high temperature. The low temperature magnetoresistance shows weak anti-localization at low field and a parabolic background in the longitudinal resistance (Fig.S6a), indicative of ordinary magnetoresistance[17]. A slight non-linearity in retrieved in the Hall coefficient (Fig.S6b), indicating the presence of multiple carrier species.

The origin of the carriers was further explored by examining the Shubnikov de-Haas (SdH) oscillations within the magnetoresistence[18]. To isolate the oscillations, the data was first smoothed with a Savitzky–Golay filter, using a 3$^{rd}$ order polynomial fit, and taking the 2nd derivative to remove any smoothly varying background. The results of this procedure are shown in Figure S6c. Both samples show oscillations above 5 T (< 0.2 T$^{-1}$) that are periodic in inverse magnetic field, with the same periodicity of 46±1 T, corresponding to a carrier density of $(2.2 \pm 0.1)\times 10^{12}$ cm$^{-2}$. The oscillations retrieved above 0.3 T$^{-1}$ are not periodic and may, in fact, arise from universal conductance fluctuations. We then assign each periodic peak in Fig.S6c, i.e. troughs in the magneto-resistance, to filled Landau levels, and construct the Landau Fan diagram shown in Fig.S6d. Due to the Berry curvature, carriers orbiting around a Dirac cone, such as those that occupy topological surface states, will pick up an extra π phase when compared to conventional carriers[18]. As such the y-intercept of a Landau fan diagram will show whether an oscillation arises from topologically trivial (an intercept of ±1 or 0) or non-trivial (±0.5) carriers. We find that, while both sets of SdH oscillations arise from a similar carrier density, the intercept of the Landau fan diagram for sample H1 is 0±0.3, whereas the intercept for sample H2 is 0.6±0.2, indicating that the oscillations in the former probably arise from topologically trivial transport, possibly due to band bending at the TI-vacuum interface, while the carriers in sample H2 are topologically protected.

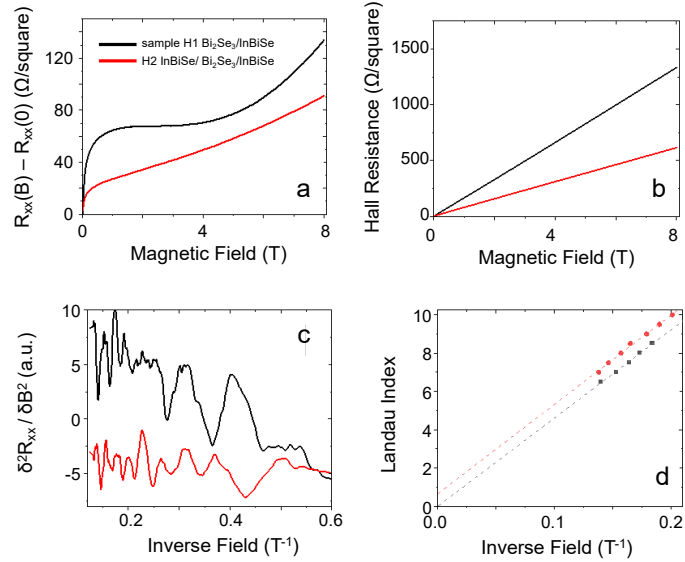

**Figure S7**: **(a-b)** (a) Magnetoresistance and (b) Hall resistance of samples H1 and H2; **(c)** 2$^{nd}$-order differential of the data in (a), plotted as a function of inverse field, showing the presence of SdH oscillations. **(d)** Landau fan diagram constructed from the data in (c), of sample H1 (black) and H2 (red).

## S10. Transmittance of the bare resonator-array transmittance

Figure S7 shows the transmittances measured on the single SRR array (SSRR, black) and the double SRR (DSRR, red) realized on bare sapphire substrate. The transmittance are measured by mounting the samples in the internal compartment of the FTIR, probed with the internal source of an FTIR spectrometer (Globar), under vacuum, in rapid scan mode (spectral resolution 1 cm$^{-1}$), and placing a wire-grid polarized in front of the array so to select the polarization-active direction and the not active polarization at 90 degrees, where the frequency transmitted is flat. The SSRR shows a resonance $\nu_{SSRR} \sim 3.17$ THz and a Q-factor $Q_{SSRR} \sim 5$, comparable with the BiSe integrated array. For the DSRR, the corresponding values are $\nu_{DSRR} \sim 3.14$ THz and $Q_{DSRR} \sim 3.2$.

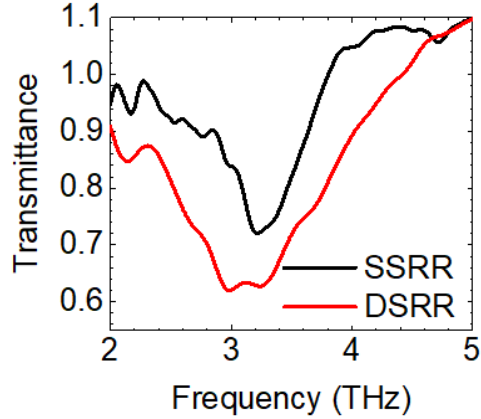

**Figure S8:** FTIR transmittance measured on the SSRR (black) and on the DSRR (red), extracted normalizing the sample transmitted signal measured when the array was illuminated by a broadband source (Globar), after filtering the linear polarization parallel to the ring dipole, with the signal acquired with the polarizer filter set at 90°.

**Section S11. Dependence of the nonlinear up-converted signal on the pumping power**

We measure the total signal detected by the Si bolometer, after filtering out the QCL fundamental mode with a 1-mm thick Ta filter, and while keeping the spectrometer moving mirror at a fixed position, on the sample A (Fig.S8a) and H1 (Fig.S8b), as a function of the power of the QCL, varied by changing the driving current, in the reasonable range in which a detectable signal can be retrieved (1.8-2.5 W). The results agree with a polynomial fit accounting for a purely 3$^{rd}$ order power law (S8a), for the sample A, while contains both a 2$^{nd}$ and 3$^{rd}$ order power law in panel b, retrieved on sample H1(S8b). In this latter case, from the fitting procedure with the function $y = 0 + Ax^2 + Bx^3$, we extract a ratio A/B = 1.25±0.37, in agreement with the experimental ratio 1.296 (see Table 1 of the main text) between the SHG and THG CEs that we have found experimentally.

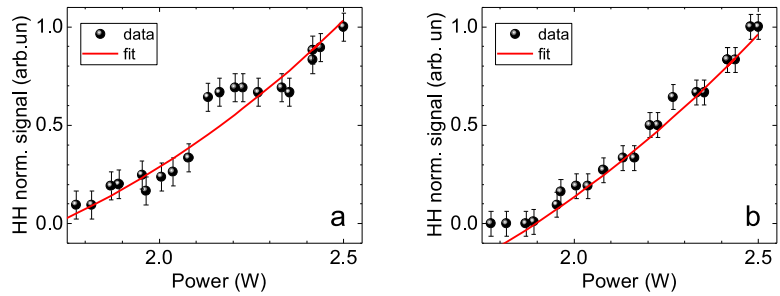

**Figure S9:** Signal measured on sample A (a) and H1(b) by the Si-bolometer with the experimental configuration of fig.3a of the main article, while keeping the FTIR moving mirror at a fixed position, as a function of QCL pump power, after filtering the QCL fundamental line with a 1mm thick Ta-filter (black dots). Fit procedure on the experimental data, perfomed with a 3$^{rd}$ order polynomial curve in **(a)**: $y = o + ax^3$, with o = -0.48 ± 0.07, a = 0.098±0.007; in **(b)**: $y = O + Ax^2 + Bx^3$, with O = -0.87 ± 0.6, A= 0.10±0.36 and B = 0.079±0.011 (red lines).

## S12. Atomic force microscope on the VdW heterostructure

The atomic force microscope images measured on sample H1 and H2 reveals mean roughness of 1.75 nm on sample H1 and 2.16 nm on sample H2.

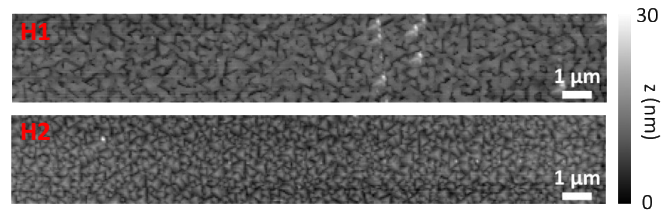

**Figure S10**: Topography of the InBiSe/BiSe heterostructure TI films, sample H1 (top) and H2 (bottom). From the image analysis, we extract a mean roughness of 1.75 nm on sample H1 and 2.16 nm on sample H2.

**References**

1. Pistore, V. *et al.* Terahertz Plasmon Polaritons in Large Area Bi2Se3 Topological Insulators. *Adv. Opt. Mater.* **n/a**, 2301673 (2023).

2. Pogna, E. A. A. *et al.* Mapping propagation of collective modes in Bi2Se3 and Bi2Te2.2Se0.8 topological insulators by near-field terahertz nanoscopy. *Nat. Commun.* **12**, 6672 (2021).

3. Di Pietro, P. *et al.* Terahertz Tuning of Dirac Plasmons in Bi2Se3 Topological Insulator. *Phys. Rev. Lett.* **124**, (2020).

4. Tielrooij, K.-J. *et al.* Milliwatt terahertz harmonic generation from topological insulator metamaterials. *Light Sci. Appl.* **11**, 315 (2022).

5. Hafez, H. A. *et al.* Terahertz Nonlinear Optics of Graphene: From Saturable Absorption to High-Harmonics Generation. *Adv. Opt. Mater.* **8**, 1900771 (2020).

6. Han, J. W. *et al.* Plasmonic Terahertz Nonlinearity in Graphene Disks. *Adv. Photonics Res.* **3**, 2100218 (2022).

7. Mics, Z. *et al.* Thermodynamic picture of ultrafast charge transport in graphene. *Nat. Commun.* **6**, 7655 (2015).

8. Tomadin, A. *et al.* The ultrafast dynamics and conductivity of photoexcited graphene at different Fermi energies. *Sci. Adv.* **4**, eaar5313 (2023).

9. Massicotte, M., Soavi, G., Principi, A. & Tielrooij, K.-J. Hot carriers in graphene – fundamentals and applications. *Nanoscale* **13**, 8376–8411 (2021).

10. Di Gaspare, A. *et al.* Electrically Tunable Nonlinearity at 3.2 Terahertz in Single-Layer Graphene. *ACS Photonics* (2023) doi:10.1021/acsphotonics.3c00543.

11. Sobota, J. A. *et al.* Ultrafast electron dynamics in the topological insulator $Bi_2Se_3$ studied by time-resolved photoemission spectroscopy. *J. Electron Spectros. Relat. Phenomena* **195**, 249–257 (2014).

12. Cox, J. D., Marini, A. & de Abajo, F. J. G. Plasmon-assisted high-harmonic generation in graphene. *Nat. Commun.* **8**, 14380 (2017).

13. Jadidi, M. M. *et al.* Tunable Terahertz Hybrid Metal–Graphene Plasmons. *Nano Lett.* **15**, 7099–7104 (2015).

14. Wang, Y. & Law, S. Optical properties of $(Bi_{1-x}In_x)_2Se_3$ thin films. *Opt. Mater. Express* **8**, 2570–2578 (2018).

15. Chudpooti, N. *et al.* Wideband dielectric properties of silicon and glass substrates for terahertz integrated circuits and microsystems. *Mater. Res. Express* **8**, (2021).

16. Richter, W., Kohler, H. & Becker, C. A Raman and Far-Infrared Investigation of Phonons. *Phys. Stat. sol* **84**, 619 (1977).

17. Spirito, D. *et al.* Weak antilocalization and spin-orbit interaction in a two-dimensional electron gas. *Phys. Rev. B* **85**, 235314 (2012).

18. Zhang, Y., Tan, Y.-W., Stormer, H. L. & Kim, P. Experimental observation of the quantum Hall effect and Berry's phase in graphene. *Nature* **438**, 201–204 (2005).